\documentclass{LMCS}

\usepackage{amsmath}
\usepackage{amssymb}
\usepackage{proof}
\usepackage{pst-all}
\usepackage{hyperref}

\newtheorem{definition}{Definition}

\newtheorem{example}{Example}

\newcommand{\comment}[1]{}
\newcommand{\comm}[1]{}

\newcommand{\nb}{\widetilde{n}}
\newcommand{\mb}{\widetilde{m}}
\newcommand{\fv}{\operatorname{fv}}
\newcommand{\fn}{\operatorname{fn}}
\newcommand{\bv}{\operatorname{bv}}
\newcommand{\bn}{\operatorname{bn}}
\newcommand{\dom}{\operatorname{dom}}
\newcommand{\ran}{\operatorname{ran}}

\newcommand{\var}{\mathcal{V}}

\newcommand{\out}[2]{\overline{ #1 }\langle #2 \rangle}
\newcommand{\bb}[1]{\overline{#1}}
\newcommand{\set}[1]{\lbrace{#1}\rbrace}
\newcommand{\subst}[2]{{}^{#2}\!/_{\!#1}}
\newcommand{\vsubst}[2]{{}^{#2}\!/_{\!#1}}

\newcommand{\lab}{\stackrel{\alpha}{\rightarrow}}         
\newcommand{\nf}[1]{{{#1}\!\!\downarrow}}

\newcommand{\nfp}{\operatorname{par}}
\newcommand{\nfpinv}{\operatorname{par}^{-1}}

\newcommand{\z}{\operatorname{z}_0}
\newcommand{\pos}{\operatorname{Pos}}
\newcommand{\posv}{\operatorname{Pos}_{\operatorname{v}}}

\newcommand{\regle}[1]{{{\scriptsize #1}}}
\newcommand{\pregle}[1]{{{\scriptsize #1}}}

\newcommand{\pair}[2]{\langle{#1},{#2}\rangle}
\newcommand{\enc}{\mathsf{enc}}
\newcommand{\dec}{\mathsf{dec}}
\newcommand{\enca}{\mathsf{enca}}
\newcommand{\deca}{\mathsf{deca}}
\newcommand{\encg}{\mathsf{enc_g}}
\newcommand{\decg}{\mathsf{dec_g}}
\newcommand{\sign}{\mathsf{sign}}
\newcommand{\scheck}{\mathsf{check}}
\newcommand{\retr}{\mathsf{retrieve}}
\newcommand{\pub}{\mathsf{pub}}
\newcommand{\priv}{\mathsf{priv}}

\newcommand{\ok}{\mathsf{ok}}
\newcommand{\head}[1]{h_{#1}}

\newcommand{\x}{\mathtt{s}}
\newcommand{\y}{\mathtt{x}}

\newcommand{\penc}[2]{\{#1\}_{#2}}       

\newcommand{\posnv}{\operatorname{Pos}_{\operatorname{nv}}}

\newcommand{\syntactic}{syntactic }
\newcommand{\Syntactic}{Syntactic }
\newcommand{\syntactically}{syntactically }
\newcommand{\simplify}{\mathsf{prune}}

\newcommand{\pdot}{\cdot}
\newcommand{\NS}{\mathit{NS}}
\newcommand{\WMF}{\mathit{WMF}}

\def\doi{3 (3:2) 2007}
\lmcsheading%
{\doi}
{1--29}
{}
{}
{Jan.~\phantom{0}5, 2007}
{Jul.~\phantom{0}6, 2007}
{}   

\begin{document}

%
%
%

 \title{Relating two standard notions of secrecy}
 \author[V.~Cortier]{V\'eronique Cortier\rsuper a}
 \address{{\lsuper a}Loria UMR 7503 \& CNRS, France}
\vskip-6 pt
 \author[M.~Rusinowitch]{Micha\"el Rusinowitch\rsuper b}
 \address{{\lsuper b}Loria UMR 7503 \& INRIA Lorraine, France}
\vskip-6 pt
 \author[E.~Z\u alinescu]{Eugen Z\u{a}linescu\rsuper c}
 \address{{\lsuper c}Loria UMR 7503  \& Universit\'{e} Henri Poincar\'{e}, France}
 \email{\{cortier,rusi,zalinesc\}@loria.fr}

 \thanks{This work has been partially supported by the ACI-SI Satin and the ACI
 Jeunes    Chercheurs JC9005.}




\maketitle

\begin{abstract}
Two styles of definitions are usually considered to express that a
security protocol preserves the confidentiality of a data $\x$.
Reachability-based secrecy means that $\x$ should never be disclosed
while equivalence-based secrecy states that two executions of a protocol with
distinct instances for $\x$ should be indistinguishable to an
attacker. Although the second formulation ensures a higher level of
security and is closer to cryptographic notions of secrecy,
decidability results and automatic tools have mainly focused on
the first definition so far.

This paper initiates a systematic investigation of the situations where \syntactic  secrecy entails
strong secrecy.
We show that in the passive case, reachability-based secrecy actually implies
equivalence-based secrecy for digital signatures, symmetric and asymmetric encryption provided
that the primitives are probabilistic.
For active adversaries,
we provide
sufficient (and rather tight) conditions on the protocol for this implication to hold.
\end{abstract}




\section{Introduction}

Cryptographic protocols are small programs designed to ensure secure communications. Since they are widely
distributed in critical systems, their security is primordial. In particular, verification using formal
methods attracted a lot of attention during this last decade. A first difficulty is to formally express the
security properties that are expected. Even a basic property such as confidentiality  admits two different
acceptable definitions namely reachability-based (\emph{syntactic}) secrecy and equivalence-based
(\emph{strong}) secrecy. Syntactic secrecy is quite appealing: it says that the secret is never accessible
to the adversary. For example, consider the following protocol where the agent $A$ simply sends a secret $s$
to an agent $B$, encrypted with $B$'s public key.
\[A\rightarrow B : \{\x\}_{\pub(B)}\]
An intruder cannot deduce $\x$, thus  $\x$ is syntactically secret. Although this notion of secrecy may be
sufficient
in many scenarios, in others, stronger security requirements are desirable.
For instance consider a setting where $\x$ is a vote and $B$ behaves differently depending on its value. If
the actions of $B$ are observable, $\x$  remains \syntactically secret but an attacker can learn the values
of the
vote by watching $B$'s actions.
The design of equivalence-based secrecy is targeted at such scenarios and intuitively says that an adversary
cannot observe the difference when the value of the secret changes. This definition is essential to express
properties like confidentiality of a vote,  of a password, or the anonymity of participants to a protocol.

Although the second formulation ensures a higher level of security and is closer to cryptographic notions of
secrecy, so far decidability results and automatic tools have mainly focused on the first definition.
The \syntactic secrecy preservation problem is undecidable in general~\cite{durgin99undecidability}, it is
co-NP-complete for a bounded number of sessions~\cite{RusinowitchT-TCS03}, and several decidable classes
have been identified in the case of an unbounded number of
sessions~\cite{durgin99undecidability,rta03,blanchetTag03,RS03}. These results often come with
automated tools,
we mention for example ProVerif~\cite{blanchet01}, Casper~\cite{lowe97casper},
CAPSL~\cite{DenkerMillenRuess00},  and Avispa~\cite{avispa-homepage-short}.

Many works have been dedicated to  proving correctness properties of protocols such as strong secrecy
using  contextual  equivalences on process calculi, like the spi-calculus.
In particular \emph{framed bisimilarity} has been introduced by Abadi and Gordon~\cite{abadi97calculus}
for this purpose. However it was not well suited for automation, as the definition of framed
bisimilarity uses several levels of quantification over infinite domains (e.g. set of contexts).
In ~\cite{elkjaer99} the authors introduce \emph{fenced bisimilarity} as an attempt to eliminate
one of the quantifiers. Also in~\cite{SymbolicBisBBNConcur04}, Borgstr\"om \textit{et al}
propose a sound but incomplete decision procedure based on a symbolic bisimulation.
Another approach to circumvent the context quantification problems is presented
in ~\cite{boreale99proof} where  labelled transition systems are  constrained
by the knowledge the environment has of names and keys. This approach allows for more direct proofs of
equivalence.
In order to get some support for compositional reasoning in this setting, \cite{boreale02compositional}
extends it with some equational laws. In  ~\cite{DuranteSV00} model-checking  techniques for the verification
of spi-calculus testing equivalence are explored. The technique is limited to finite processes but seems
to perform well on some examples. The concept of 
logical relations for the polymorphic lambda calculus has also 
been been employed to prove behavioral equivalences between programs that rely on encryption in a compositional manner~\cite{SumiiP03}. 

However, to the best of our knowledge, the only tool capable of verifying strong
secrecy is the resolution-based algorithm of ProVerif~\cite{BlanchetOakland04} that has been extended for
this purpose.
Proverif has also been enhanced for handling equivalences of processes that differ only in the choice of
some
terms  in the context of the applied pi calculus \cite{BlanchetAbadiFournetLICS05}. This allows to add
some equational theories for modelling properties of the underlying cryptographic primitives.

Similarly very few decidability results are available for strong secrecy. In the article~\cite{huttel},
H\"uttel proves
decidability for a fragment of the spi-calculus without recursion for framed bisimilarity.
For recursive processes only a class of ping-pong protocols restricted to two principals admits
a decidable strong bisimilarity relation~\cite{HS05}.

Finally, we should mention here some related works based on the concept of non-interference~\cite{ryan99process}.  
This notion formalizes the absence of unauthorized information flow in multilevel computer systems. 
Non-interference has been widely investigated in the context of langage-based security (e.g.~\cite{volpano,zdancewic01robust}).
It can be expressed  with process equivalence techniques and has been applied 
also to security protocols in \cite{focardi00non,BugliesiCR03}. 
An advantage of this approach is that various security properties, including secrecy, 
can be modeled by selecting  proper equivalence relations. However as far as we know 
decidability results for non-interference  properties of security protocols have not been reported.

In light of the above discussion, it may seem that the two notions of secrecy are separated by a sizable gap
from both a conceptual but also from a practical point of view. These two notions have counterparts in the
cryptographic setting (where messages are bitstrings and the adversary is any polynomial probabilistic
Turing machine). Intuitively, the \syntactic secrecy notion can be translated into a similar
reachability-based
secrecy notion and equivalence-based notion is close to indistinguishability. A quite surprising
result~\cite{esop05} states that cryptographic \syntactic secrecy actually implies indistinguishability in
the
cryptographic setting. 
This result relies in particular on the fact that the encryption schemes are probabilistic thus two
encryptions of the same plaintext lead to different ciphertexts.

Motivated by the result of~\cite{esop05} and the large number of available systems for \syntactic secrecy
verification, we initiate in this paper a systematic investigation of situations where \syntactic secrecy
entails
strong secrecy. Surprisingly, this happens in many interesting cases.

We offer results in both passive and active cases in the setting of the \emph{applied pi
cal\nolinebreak culus}~\cite{AbadiFournet01}. We first treat in Section~\ref{sec:passive} the case of passive
adversaries. We prove that \syntactic secrecy is equivalent to  strong secrecy. This holds for signatures,
symmetric and asymmetric encryption. It can be easily seen that the two notions of secrecy are not
equivalent in the case of deterministic encryption. Indeed, the secret $\x$ cannot be deduced from the
encrypted message $\{\x\}_{\pub(B)}$ but if the encryption is deterministic, an intruder may try different
values for $\x$  and check whether the ciphertext he obtained using $B$'s public key is equal to the one he
receives. Thus for our result to hold, we require that encryption is probabilistic. This is not a
restriction since this is \textit{de facto} the standard in almost all cryptographic applications.
Next, we consider the more challenging case of active adversaries. We give sufficient conditions on the
protocols for \syntactic secrecy to imply strong secrecy (Section~\ref{sec:active}). Intuitively, we require
that
the conditional tests are not performed directly on the secret since we have seen above that such tests
provide information on the value of this secret. We again exhibit several counter-examples to motivate the
introduction of our conditions. An important aspect of our result is that we do not make any assumption on
the number of sessions: we put  no restriction on the use of
replication.
In particular, our result holds for an unbounded number of sessions.

The interest of our contribution is twofold. First, conceptually, it helps  to understand when the two
definitions of secrecy are actually equivalent. Second, 
we can transfer many  existing results (and the armada of automatic
tools) developed for \syntactic secrecy. For instance, since the
\syntactic secrecy problem is decidable for tagged protocols  for an
unbounded number of sessions~\cite{RS03}, by translating the tagging
assumption to the applied-pi calculus, we can derive a first
decidability result for strong secrecy for an unbounded number of
sessions. Other decidable fragments might be derived
from~\cite{durgin99undecidability} for bounded messages (and nonces)
and~\cite{AL00} for a bounded number of sessions.
A first version of this result was published in the Proceedings
of CSL'06~\cite{CortierRZ-CSL06}, with no detailed proofs. In that preliminary
version, the correspondence result in the active case was only
established for symmetric encryption. We extend it here to asymmetric
encryption and digital signatures.




\section{Passive case}\label{sec:passive}
\subsection{Syntax}

Cryptographic primitives are represented by function symbols. More specifically, we consider the signature
$\Sigma = \{\enc$, $\dec$, $\enca$, $\deca$, $\pub$, $\priv$, $\langle\rangle$, $\pi_1$, $\pi_2$, $\sign$,
$\scheck$, $\retr\}$ where the function symbols have arities $3,2,3,2,1,1,2,1,1,2,3$ and $1$ respectively.
$\mathcal{T}(\Sigma,\mathcal{X},\mathcal{N})$, or simply $\mathcal{T}$, denotes the
set of terms built over $\Sigma$ extended by a  set of constants, the infinite set of \emph{names}
$\mathcal{N}$ and the infinite set of variables $\mathcal{X}$. A term is \emph{closed} or \emph{ground} if
it does not contain any variable. The set of names occurring in a term $T$ is denoted by $\fn(T)$, the set
of variables is denoted by $\var(T)$. The \emph{positions} in a term $T$ are defined recursively as usual
(\textit{i.e.} as sequences of positive integers), $\epsilon$ being the empty sequence. Denote by
$\mathbb{N}^*_+$ the set of sequences of positive integers. We denote by $T|_{p}$ the  subterm of $T$ at
position $p$ and by $U[V]_p$ the term obtained by replacing in $U$ the subterm at position $p$ by
$V$. $\pos(T)$ denotes the set of positions of
$T$, $\posv(T)$ the set of positions of variables in $T$ and $\posnv(T)=\set{p\in \pos(T)\mid
T|_p\notin\var(T)}$ the set of non-variable positions of $T$.  We
may simply say that a term $V$ \emph{is in} a term $U$ if $V$ is a
subterm of $U$
We denote by $\le_{st}$
(resp.~$ <_{st}$) the subterm (resp. strict) order. $\head{U}$ denotes the function
symbol, name or variable at position $\epsilon$ in the term~$U$.
A substitution is a function that maps variables to terms
$\sigma:\mathcal{X}\rightarrow \mathcal{T}$.
We write $\sigma = \{\subst{x_1}{T_1},\ldots\subst{x_n}{T_n}\}$ to say
that $x_i\sigma = T_i$ for $1\leq i\leq n$ and $x\sigma = x$ for
$x\neq x_i$. The expression $U[\subst{x}{V}]$ denotes $U\sigma$ where
$\sigma  = \{\subst{x}{V}\}$. 

We equip the signature with an equational theory~$E$:
\[\left\{\begin{array}{l}
\pi_1(\langle z_1,z_2\rangle)=z_1\\
\pi_2(\langle z_1,z_2\rangle)=z_2\\
\dec(\enc(z_1,z_2,z_3),z_2)=z_1\\
\deca(\enca(z_1,\pub(z_2),z_3),\priv(z_2))=z_1\\
\scheck(z_1,\sign(z_1,\priv(z_2)),\pub(z_2))=\ok\\
\retr(\sign(z_1,z_2))=z_1\\
\end{array}\right.\]
Let $\mathcal{R}_E$ be the corresponding rewrite system (obtained by orienting the equations from left to
right). $\mathcal{R}_E$ is convergent. The normal form of a term $T$ w.r.t.~$\mathcal{R}_E$ is denoted by
$\nf{T}$. Notice that $E$ is also stable by substitution of names. As usual, we write $U\rightarrow V$ if
there exists $\theta$, a position $p$ in $U$ and $L\rightarrow R\in \mathcal{R}_E$ such that $U|_p=L\theta$
and $V=U[R\theta]_p$.

The symbol $\langle\_,\_\rangle$ represents the pairing function and $\pi_1$ and $\pi_2$ are the associated
projection functions. The term $\enc(M,K,R)$ represents the message $M$ encrypted with the key $K$. The
third argument $R$ reflects that the encryption is probabilistic: two encryptions of the same messages under
the same keys are different. The symbol $\dec$ stands for decryption. The symbols $\enca$ and $\deca$ are
very similar but in an asymmetric setting, where $\pub(a)$ and $\priv(a)$ represent respectively the public
and private keys of an agent $a$. We denote by $\encg$
(respectively $\decg$) a generic encryption (decryption), that is when using it we refer to both symmteric
\emph{and} asymmetric encryption (decryption). The term $\sign(M,K)$ represents the signature of  message $M$
with  key $K$. $\scheck$ enables to verify the signature and $\retr$ enables to retrieve the signed message
from the signature.\footnote{Signature schemes may disclose partial information
  on the signed message. To enforce the intruder capabilities, we
  assume that messages can always be retrieved out of
  the signature.} The function symbols $\langle\rangle, \enc, \enca$ and $\sign$ are called
\emph{constructors}, while $\pi_1, \pi_2, \dec, \deca, \scheck$ and $ \retr$ are called \emph{destructors}.

After the execution of a protocol, an attacker knows the messages sent on the network and also in which
order they  were sent. Such message sequences  are organized as \emph{frames} $\varphi=\nu\nb.\sigma$, where
$\sigma=\set{\subst{y_1}{M_1},\dots,\subst{y_l}{M_l}}$ is an acyclic 
substitution and $\nb$ is a finite
set of
names. We denote $\dom(\varphi)=\dom(\sigma)=\{y_1,\dots,y_l\}$ and
$\ran(\varphi)=\ran(\sigma)=\{M_1,\dots,M_l\}$. The variables $y_i$ enable us to refer
to each message. The names in $\nb$ are said to be \emph{restricted} in $\varphi$. Intuitively, these names
are \textit{a priori} unknown to the intruder. The names outside $\nb$ are said to be \emph{free}  in
$\varphi$. 
The set of free names occurring in $\varphi$ is denoted $\fn(\varphi)$.
A term
$M$ is said \emph{public} w.r.t.~a frame $\nu\nb.\sigma$ (or w.r.t.~a set of names $\nb$) if $\fn(M)\cap
\nb=\emptyset $ and it does not use the function symbol $\priv$; in other words if
$M\in\mathcal{T}(\Sigma\!\setminus\!\{\priv\},\mathcal{X},\mathcal{N}\!\setminus\!\nb)$. The frame
or the set of names might be omitted when it is clear from the context. We usually write $\nu n_1,\ldots,n_k$
instead of $\nu \set{n_1,\ldots,n_k}$.

\subsection{Deducibility}\label{sec:deducibility}
Given a frame $\varphi$ that represents the history of messages sent during the execution of a protocol, we
define the \emph{deduction} relation, denoted by $\varphi\vdash M$. Deducible messages are messages that can
be obtained from $\varphi$ by applying function symbols and the equational theory $E$.
\[
\begin{tabular}{c@{\quad\quad\quad}c}
\infer[x\in\dom(\sigma)]{\nu\nb.\sigma\vdash x\sigma}{} &
\infer[ m\in\mathcal{N}\backslash\nb]{\nu\nb.\sigma\vdash m}{}\\\\
\infer[f\neq\priv]{\nu\nb.\sigma\vdash f(T_1,\dots,T_l)}{\nu\nb.\sigma\vdash T_1 & \cdots &
\nu\nb.\sigma\vdash T_l} & \infer{\nu\nb.\sigma\vdash T'}{\nu\nb.\sigma\vdash T &
T=_E T'}\\
\end{tabular}
\]

\begin{example} \label{ex:deducibility}
$k$ and $\langle k,k'\rangle$ are deducible from the frame $\nu
k,k',r.\set{\subst{x}{\enc(k,k',r)},\subst{y}{k'}}$.
\end{example}

A message is usually said secret if it is not deducible. By opposition to our next notion of secrecy, we say
that a term $M$ is \emph{\syntactically
  secret} in $\varphi$ if $\varphi\not\vdash M$.

We will often use another characterization of deducible terms.
\begin{prop}\label{prop_deduc}
Let $\varphi = \nu\nb.\sigma$ be a frame and $M$ be a term. $\varphi\vdash M$ if and
only if there exists a public term $T$ w.r.t.~$\varphi$ such that
$T\sigma = _E M$.
\end{prop}
This is easily proved by induction on the length of the proof of deducibility.

\subsection{Static equivalence}
Deducibility does not always suffice to express the abilities of an intruder.
\begin{example}\label{ex:notstatic}
The set of deducible messages is the same for the frames 
$\varphi_1 = \nu
k,n_1,n_2,r_1.\\\{\subst{x}{\enc(n_1,k,r_1)},\subst{y}{\langle
    n_1,n_2\rangle}, \subst{z}{k}\}$
and $\varphi_2 = \nu k,n_1,n_2,r_1.\set{\subst{x}{\enc(n_2,k,r_2)},\subst{y}{\langle
    n_1,n_2\rangle},\subst{z}{k}}$, while an
attacker is able to detect that the first message corresponds to distinct nonces. In particular, the attacker
is able to distinguish the two ``worlds'' represented by $\varphi_1$ and $\varphi_2$.
\end{example}

We say that a frame $\varphi=\nu\nb.\sigma$ \emph{passes the test}
$(U,V)$ where $U,V$ are two terms, denoted by $(U=V)\varphi$, if there
exists a renaming of the restricted names in$\varphi$ such that
$(\fn(U)\cup\fn(V))\cap \nb =\emptyset$ and $U\sigma=_E V\sigma$. Two
frames $\varphi=\nu\nb.\sigma$ and $\varphi'=\nu\mb.\sigma'$ are
\emph{statically equivalent}, written $\varphi\approx\varphi'$, if
they pass the same public tests, that is, if
$\dom(\varphi)=\dom(\varphi')$ and for all public terms $U,V$
w.r.t.~$\varphi$ and $\varphi'$ such that
$(\var(U)\cup\var(V))\subseteq \dom(\varphi)$ we have $(U= V)\varphi$
if and only if $(U= V)\varphi'$.

\begin{example} The frames $\varphi_1$ and $\varphi_2$ defined in
  Example~\ref{ex:notstatic} are not statically equivalent
since $(\dec(x,z)=\pi_1(y))\varphi_1$ but $(\dec(x,z)\neq \pi_1(y))\varphi_2$.
\end{example}

\comment{ Of course an intended syntactical secret name $\x$ must be restricted, but when talking about
instances of $\x$ we must consider it (at least) a free name (if not a variable). Hence we compare
\syntactic secrecy and strong secrecy regarding the same frame modulo the restriction on the secret $\x$. We
use the notation $\nu\x.\varphi$ for $\nu(\nb\cup\set{\x}).\sigma$, where $\varphi=\nu\nb.\sigma$. Thus $\x$
is syntactically secret if $\nu\x.\varphi\nvdash\x$.}

Let $\varphi = \nu\nb.\sigma$ be a frame and $\x\in\nb$ a restricted name in $\varphi$. Let $M$ be a term
such that $\fn(M)\cap\nb=\emptyset$. We denote by $\varphi[\subst{\x}{M}]$ the frame
$\nu\nb.\sigma[\subst{\x}{M}]$ obtained by instantiating $\x$ with $M$ in each term of the
substitution $\sigma$.

We say
that $\x$ is \emph{strongly secret} in $\varphi$ if for every closed
public terms $M, M'$ w.r.t.~$\varphi$, we have
$\varphi[\subst{\x}{M}]\approx\varphi[\subst{\x}{M'}]$ that is, the
intruder cannot distinguish the frames obtained by  instantiating the secret $\x$ by two terms of
its choice. For simplicity we may omit  $\x$ and write $\varphi[M]$
instead of $\varphi[\subst{\x}{M}]$.

\subsection{\Syntactic secrecy implies strong secrecy}\label{sec:implic}
\Syntactic secrecy is usually weaker than strong secrecy! We first
exhibit some examples of frames that preserves \syntactic secrecy
but not strong secrecy. They all rely on different properties.
\smallskip

{\bf Probabilistic encryption.} The frame $\psi_1 = \nu
\x,k,r.\set{\subst{x}{\enc(\x,k,r)},\subst{y}{\enc(n,k,r)}}$  does not
  preserve the strong secrecy of $\x$.
Indeed, $\psi_1[n]\not\approx \psi_1[n']$ since ${(x=y)}\,\psi_1[n]$ but ${(x\neq y)}\,\psi_1[n']$. This
would not
happen if each encryption used a distinct randomness,
  that is if the encryption was probabilistic.
\smallskip

{\bf Key position.} The frame $\psi_2\!=\!\nu\x,\!n.\{\subst{x}{\enc(\langle n,n'\rangle,\x,r)}\}$  does
not  preserve the strong secre\-cy of $\x$.
Indeed, $\psi_2[k]\not\approx \psi_2[k']$ since ${(\pi_2(\dec(x,k)) =
  n')}\,\psi_2[k]$ but 
${(\pi_2(\dec(x,k)) \neq
n')}\,\psi_2[k']$. If $\x$ occurs in key position in some ciphertext, the intruder may try to decrypt
the ciphertext since $\x$ is replaced by public terms and check for some redundancy. It may occur that the
encrypted message does not contain any
  verifiable part. In that case, the frame may preserve strong
  secrecy. It is for example the case for the frame $\nu
  n.\!\set{\subst{x}{\enc(n,\x,r)}}$. Such cases are however quite rare
  in practice.
\smallskip

{\bf No destructors.} The frame $\psi_3 =
\nu\x.\set{\subst{x}{\pi_1(\x)}}$ does not preserve the strong secrecy of
$\x$ simply because $(x=k)$ is true for $\psi_3[\langle
k,k'\rangle]$ while not for $\psi_3[k]$.
\smallskip

{\bf Retrieve rule.} The $\retr(\sign(z_1,z_2))\!=\!z_1$ equation may seem
arbitrary since not all signature schemes enable to get the signed
message out of a signature. It is actually crucial for our result.
For example, the frame $\psi_4 = \nu\x.\set{\subst{x}{\sign(\x,\priv(a))},
\subst{y}{\pub(a)}}$ does not preserve the strong secrecy of $\x$
because $(\scheck(n,x,y)=\ok)$ is true for $\psi_4[n]$ but not for
$\psi_4[n']$.

In the three first cases, the frames preserve the \syntactic secrecy of
$\x$, that is $\psi_i\not\vdash \x$, for $1\leq i\leq 3$.
In the fourth case, we would also have $\psi_4\not\vdash \x$ without
the $\retr$ equation.
\smallskip

We define agent encryptions as encryptions which  use ``true'' randomness, that is fresh
names. Note that in the passive case all encryptions are produced by agents and not by the intruder.
Encryption (as a primitive) is probabilistic if each (instance of the) encryption uses a distinct
randomness. Next, we define those notions formally.

We say that an occurrence $q_{\enc}$
of an encryption in a term $U$ is an \emph{agent encryption} {w.r.t.} a set of names $\nb$ if
$U|_{q_{\enc}\pdot 3}\in\nb$.
We say that an occurrence $q_{\enc}$ of an
encryption in a term $U$ is a \emph{probabilistic encryption} {w.r.t.}
a set of terms $S$ if no distinct term shares the same randomness,
that is, for any term
$V\in S$ and position $p$ such that $V|_p=U|_{q_{\enc}\pdot 3}$ we have that $p=q\pdot 3$ for some $q$ and
$V|_q=U|_{q_{\enc}}$.

\comment{We say that a set $\mathcal{S}$ of terms has the
\emph{``unique randomness''} property w.r.t.~a set of names $\nb$ if for any agent encryption $q_{\enc}$ in
some $T\in S$ and any term $T'\in \mathcal{S}$ and position $p$ such that $T'|_p=T|_{q_{\enc}\pdot 3}$ we
have that
$p=q\pdot 3$ for some $q$ and $T'|_q=T|_{q_{\enc}}$.}

\newcommand{\defgood}{
A frame $\varphi=\nu\nb.\sigma$ is \emph{well-formed}
\textit{w.r.t.} some name $\x$ if
\begin{enumerate}
\comment{\item \label{cond_nondet} Any encryption in $\sigma$ is probabilistic w.r.t.~the set of terms of
$\sigma$, and for any occurrence of $\x$ (say $q_{\x}$ in $y\sigma$) the lowest encryption above $\x$
(that is $\max\{q\mid q<q_{\x} \wedge \head{y\sigma|_q}=\encg\}$), if it exists, is an agent encryption
w.r.t.~$\nb\!\setminus\!{\set{\x}}$.}

\comment{for any subterm $\enc(M,K,R)$ of $\varphi$, for any term  $T\in \varphi$ and position $p$  such that
$T|_p=R$ we have $p=q.3$ for some $q$ and $T|_q= \enc(M,K,R)$.
In addition, if $\x$ occurs in $M$ at a position $p'$
such that no encryption appears along the path from the root to $p'$ then
 $R$ must be restricted, that is $R\in\nb$. The same conditions  hold for asymmetric encryption. }

\comment{Encryption is probabilistic,
  \textit{i.e.} all encryptions in $\sigma$ are agent encryptions and the set of terms of $\sigma$ has
``unique randomness'' property w.r.t.~$\nb$.}

\item \label{cond_nondet} any encryption in $\sigma$ is an agent encryption w.r.t.~$\nb\!\setminus\!{\set{\x}}$ and a probabilistic encryption w.r.t.~the set of terms of $\sigma$;

\item \label{cond_no_x_in_key} $\x$ is not  part of a key or a randomness, \textit{i.e.}~for all
$\enc(M,K,R)$,
$\enca(M',K',R')$, $\sign(U,V)$, $\pub(W)$, $\priv(W')$ subterms of $\varphi$,
$\x\notin\fn(K,K',V,W,W',R,R')$;
\item \label{cond_nodec} $\varphi$ does not contain destructor symbols.
\end{enumerate}
}

The previous examples lead us to the following definition.
\begin{definition}\label{def_good}
\defgood
\end{definition}

\comment{
Condition \ref{cond_nondet} requires that each innermost encryption above $\x$ contains a
restricted randomness. This  is not a restriction since  $\x$ is meant to be a secret value and  such
encryptions have to be produced by honest agents and  thus contain a restricted randomness.
}

\newcommand{\thpas}{
Let $\varphi$ be a well-formed frame \textit{w.r.t.} $\x$, where
$\x$ is a restricted name in $\varphi$.
\[
\varphi\nvdash \x \ \mbox{ if and only if }\
 \varphi[\subst{\x}{M}]\approx \varphi[\subst{\x}{M'}]
\]
for all $M, M'$ closed public terms w.r.t.~$\varphi$.
}

For well-formed frames, \syntactic secrecy is actually equivalent to strong secrecy.
\begin{thm}\label{th}
\thpas
\end{thm}

\newcommand{\lemmasyntactic}{
Let $\varphi=\nu\nb.\sigma$ be a well-formed frame w.r.t.~$\x\in\nb$ such that $\varphi\nvdash\x$. Let $U$,
$V$ and $M$ be public terms w.r.t.~$\varphi$, with $\var(U),\var(V)\subseteq\dom(\sigma)$ and $M$ ground.
Then $U\sigma[\subst{\x}{M}]=V\sigma[\subst{\x}{M}]$ implies $U\sigma=V\sigma$.
}

\newcommand{\lemmasimulate}{Let $\varphi=\nu\nb.\sigma$ be a well-formed frame w.r.t.~$\x\in\nb$
such that $\varphi\nvdash \x$. Let $U$ be a term with
$\var(U)\subseteq \dom(\varphi)$ and $M$ be a closed term in normal
form such that $U$ and $M$ are public w.r.t.~$\varphi$.
 If $U\sigma[\subst{\x}{M}]\rightarrow V$, for some term $V$, then
  there exists a frame
$\varphi'=\nu\nb.\sigma'$ well-formed w.r.t.~$\x$
\begin{itemize}
\item extending $\varphi$, that is $x\sigma'=x\sigma$ for all $
  x\in\dom(\sigma)$,
\item preserving deducible terms: $\varphi\vdash W$ if and only if
  $\varphi'\vdash W$,
\item and such that  $V=V'\sigma'[\subst{\x}{M}]$ and $U\sigma\rightarrow V'\sigma'$ for some $V'$
public w.r.t.~$\varphi'$.
\end{itemize}}
\proof
Let $\varphi = \nu\nb.\sigma$ be a well-formed frame w.r.t.~$\x$. If
$\varphi\vdash \x$, this trivially implies that $\x$ is not
strongly secret. Indeed, there exists a public term $T$ w.r.t.~$\varphi$
such that $T\sigma =_E \x$, by Proposition~\ref{prop_deduc}.
Let $n_1,n_2$ be fresh names
such that $n_1,n_2\notin\nb$ and $n_1,n_2\notin\fn(\varphi)$. Since
$T\sigma[\subst{\x}{n_1}] =_E n_1$ the frames
$\varphi[\subst{\x}{n_1}]$ and $\varphi[\subst{\x}{n_2}]$ are
distinguishable with the test $(T=n_1)$.

We assume now that $\varphi\nvdash \x$. We first show that any \syntactic equality satisfied by the frame
$\varphi[\subst{\x}{M}]$ is already satisfied by $\varphi$.
\begin{lem}\label{lemma_syntactic}
\lemmasyntactic
\end{lem}
This lemma is proved in Subsection~\ref{sec:proof:passif}.

The key lemma is that any reduction that applies to a deducible term $U$ where $\x$ is replaced by some
$M$, directly applies to $U$.
\begin{lem}\label{lemma_simulate}
\lemmasimulate
\end{lem}
This lemma (proved in Subsection~\ref{sec:proof:passif}) allows us to conclude the proof of Theorem~\ref{th}. Fix arbitrarily
two public closed terms
$M,M'$. We can assume w.l.o.g. that $M$ and $M'$ are in normal form. Let $U\neq V$ be two public terms such
that $\var(U),\var(V)\subseteq\dom(\varphi)$ and $U\sigma[\subst{\x}{M}]=_E V\sigma[\subst{\x}{M}]$. Then
there are $U_1,\dots,U_k$ and $V_1,\dots,V_l$ such that $U\sigma[\subst{\x}{M}]\!\rightarrow\!
U_1\!\rightarrow\!\dots\!\rightarrow\! U_k$, $V\sigma[\subst{\x}{M}]\!\rightarrow\!
V_1\!\rightarrow\!\dots\!\rightarrow\! V_l$, $U_k=U\sigma[\subst{\x}{M}]\!\!\downarrow$,
$V_l=V\sigma[\subst{\x}{M}]\!\!\downarrow$ and $U_k=V_l$.

Applying repeatedly Lemma~\ref{lemma_simulate} we obtain that there exist public terms $U'_1,\dots,U'_k$ and
$V'_1,\dots,V'_l$ and well-formed frames $\varphi_{i}=\nu\nb.\sigma_{i}$, for $i\in\{1,\dots, k\}$ and
$\psi_{j}=\nu\nb.\theta_{j}$, for $j\in\{1,\dots, l\}$ (as in the lemma) such that
$U_i=U'_i\sigma_{i}[\subst{\x}{M}]$, $U\sigma\rightarrow U'_1\sigma_1$, $U'_i\sigma_{i}\rightarrow
U'_{i+1}\sigma_{{i+1}}$, $V_j = V'_j\theta_{j}[\subst{\x}{M}]$, $V\sigma\rightarrow V'_1\theta_1$ and
$V'_j\theta_{j}\rightarrow V'_{j+1}\theta_{{j+1}}$.


The substitution $\sigma_{k}$ extends $\sigma$, which means that
$\sigma_{k} = \sigma\cup \sigma'_{k}$ with
$\dom(\sigma)\cap\dom(\sigma'_k) = \emptyset$.
Similarly, $\theta_l = \sigma\cup\theta'_l$ with
$\dom(\sigma)\cap\dom(\theta'_l) = \emptyset$.
By possibly renaming the variable of $\theta'_l$ and of the $V'_j$, we
can assume that $\dom(\sigma'_k)\cap\dom(\theta'_l) = \emptyset$.
We consider $\varphi'=\nu\nb.\sigma'$ where
$\sigma'=\sigma\cup\sigma'_{k}\cup\theta'_{l}$. Since only subterms of
$\varphi$ have been added to $\varphi'$, it is easy to verify that
$\varphi'$ is still a well-formed frame and for every term $W$ we have that
$\varphi\vdash W$ if and only if $\varphi'\vdash W$. {In particular
$\varphi'\nvdash\x$.}

By construction we have that
 $U'_k\sigma_{k}[\subst{\x}{M}]=V'_l\theta_{l}[\subst{\x}{M}]$.
Then, by Lemma~\ref{lemma_syntactic}, we deduce that
$U'_k\sigma_{k}=V'_l\theta_{l}$ that is
$U\sigma=_E V\sigma$. By stability of substitution of names, we have
$U\sigma[\subst{\x}{M'}]=_E V\sigma[\subst{\x}{M'}]$. We deduce
that $\varphi[\subst{\x}{M}]\approx \varphi[\subst{\x}{M'}]$.\qed

\subsection{Generalization of well-formed frames}
\label{sec:proof:passif}
In the active case, we need a more general definition for well-formed
frames and for the corresponding lemmas.
In particular, we need to consider frames with destructor symbols.
Thus we provide here the definition of \emph{extended well-formed}
frames, show that well-formed frames are special cases of extended well-formed (when the
frames preserve \syntactic secrecy), and then prove analogue lemmas for extended well-formed frames.


We say that there is an encryption \emph{plaintext-above} a subterm $T$ of a term
$U$ at position $q_T$ if there is a position $q<q_{T}$ such that $U|_q$ is  a
cyphertext, that is $h_{U|_{q}}\in \set{\enc, \enca}$. In
addition, $T$ occurs in the plaintext subterm  of the encrypted
term, that is $q\pdot 1\leq q_{T}$.

\newcommand{\defgoodrev}{
We say that a frame $\varphi=\nu\nb.\sigma$ is an \emph{extended
well-formed} w.r.t.~$\x$ if (1) all the terms of $\sigma$ are in normal form, (2) any agent encryption
w.r.t.~$\nb$ in $\sigma$ is a probabilistic encryption w.r.t.~$\ran(\sigma)$,
and (3) for every occurrence $q_{\x}$ of $\x$ in $y\sigma$ with $y\in\dom(\sigma)$, there exists an agent
encryption (say $q_{\enc}$) {w.r.t.} $\nb\!\setminus\!\!\set{\x}$ plaintext-above $\x$.
In addition, (4) the lowest agent encryption $q_0$ plaintext-above $\x$ satisfies
$\head{y\sigma|_q}\in\{\langle\rangle,\sign\}$, for all positions $q$ with $q_0<q<q_{\x}$. }
\begin{definition}\label{def_good_rev}
\defgoodrev
\end{definition}
This definition ensures in particular that there is no destructor directly above~$\x$.

\begin{example}
The frame $\varphi=\nu \x,k,n.\set{\subst{x}{\pi_1(\enc(a,\enc(\pair{b}{\x},k,n)),n'')},
\subst{y}{\enc(a,k',n')}, \subst{z}{\enc(b,k',n')}}$ is extended well-formed, while the frames
$\varphi_2=\nu n.\set{\subst{y}{\enc(a,k,n)}, \subst{z}{\enc(b,k,n)}}$, $\varphi_3=\nu
n.\set{\subst{x}{\enc(a,\x,n)}}$, and $\varphi_4=\nu \x,k,n.\set{\subst{x}{\enc(\pi_1(\x),k,n)}}$ are not,
each frame $\varphi_i$ contradicting condition $(i)$.
\end{example}

\newcommand{\lemmadeduc}{
Let $\varphi=\nu\nb.\sigma$ be a well-formed frame w.r.t.~$\x\in\nb$
and let $q_{\x}$ be an occurrence of $\x$ in $y\sigma$ for
some $y\in\dom(\sigma)$. If $\varphi\nvdash\x$ then there
is an encryption plaintext-above $s$, that is
exists a position $q<q_{\x}$ such that $y\sigma|_q$ is  a
cyphertext, that is $h_{y\sigma|_{q}}\in \set{\enc, \enca}$. In
addition, $\x$ occurs in the plaintext subterm  of the encrypted
term, that is $q\pdot 1\leq q_{\x}$.
}

We first start by a preliminary lemma which states that in a  well-formed frame w.r.t.~$\x$, either  every
occurrence of $\x$ is under some encryption or $\x$ is deducible.
\begin{lem}\label{lemmadeduc}
\lemmadeduc
\end{lem}
\proof
Assume by contradiction that there is  an occurrence of $\x$ such that there is no encryption plaintext-above
$\x$.
Then, from Properties~\ref{cond_no_x_in_key} and~\ref{cond_nodec} of well-formed frames, we have that  there
are only pairs and signatures as function symbols above $\x$. Hence $\x$ is  deducible (by applying
the projections and the $\retr$ equations). Thus there exists a position $q<q_{\x}$ such that
$y\sigma|_q$ is  an encryption. By Property~\ref{cond_no_x_in_key}  of well-formed
frames, $\x$ must occur in the plaintext part of the encryption that is $q\pdot 1\leq q_{\x}$.\qed

\begin{lem}\label{lem:well-ext}
Let $\varphi=\nu\nb.\sigma$ be a frame and $\x$ a restricted name in $\varphi$ such that $\varphi\nvdash\x$.
If $\varphi$ is a well-formed frame w.r.t.~$\x$ then it is an extended well-formed frame w.r.t.~$\x$.
\end{lem}
\proof
Since there are no destructor symbols in $\varphi$ all terms are in normal form. Since any encryption in
$\sigma$ is probabilistic it will be a fortiori the case for agent encryptions.

Consider an occurrence $q_{\x}$ of $\x$ in $y\sigma$ with $y\in\dom(\sigma)$. From Lemma~\ref{lemmadeduc}
we have that there is at least an encryption plaintext-above $\x$ in $y\sigma$. Consider the lowest one. Then
condition
\ref{cond_nondet} of well-formed frames says that this encryption is an agent encryption.
Conditions~\ref{cond_no_x_in_key} and~\ref{cond_nodec} impose that the only function symbols in between may
be $\langle\rangle$ and $\sign$.\qed


The following lemma states that if in two distinct terms the secret is protected by agent probabilistic
encryptions then by replacing the secret with any term we cannot obtain two syntactically equal terms.

\begin{lem}\label{lemma_synt}
Let $\nb$ be a set of names and  $\x$ be a name, $\x\in\nb$. Let $M$ be a ground public term w.r.t.~$\nb$ and
$U, V$ be two terms such that for any occurrence $q_{\x}$ of $\x$ (in $U$ or $V$) there is an
encryption $q_{\enc}$   (in $U$ or $V$ respectively) with $q_{\enc}\pdot 1\le q_{\x}$ such that $q_{\enc}$ is
an agent encryption w.r.t.~$\nb\!\setminus\!\!\set{\x}$ and  $q_{\enc}$ is a probabilistic encryption w.r.t.~$\set{U,V}$. Then $U[\subst{\x}{M}]=V[\subst{\x}{M}]$ implies $U=V$.
\end{lem}
\proof
Suppose that $U[\subst{\x}{M}]=V[\subst{\x}{M}]$ and $U\neq V$. Then there is an
occurrence $q_{\x}$ of $\x$, say in $U$, such that $V|_{q_{\x}}\neq\x$. Consider an agent probabilistic
encryption $q_{\enc}$ with $q_{\enc}\pdot 1\le q_{\x}$ as in the lemma. We have $U|_{q_{\enc}\pdot
3}\in\nb\!\setminus\!\!\set{\x}$. It follows that $V[\subst{\x}{M}]|_{q_{\enc}\pdot
3}\in\nb\!\setminus\!\!\set{\x}$. Since $M$ is public this implies that $q_{\enc}\pdot 3$ is a position in
$V$. And since $q_{\enc}$ is a probabilistic encryption and $U|_{q_{\enc}\pdot 3}= V|_{q_{\enc}\pdot 3}$ it
follows that $U|_{q_{\enc}}=V|_{q_{\enc}}$. Hence $U|_{q_{\x}}=V|_{q_{\x}}$ which represents a contradiction
with $V|_{q_{\x}}\neq\x$.\qed

\begin{cor}\label{corol_synt}
Let $\varphi=\nu\nb.\sigma$ be an extended well-formed frame w.r.t.~$\x\in\nb$ such that $\varphi\nvdash
\x$. Let $U$, $V$ and $M$ be public terms w.r.t.~$\varphi$, with $\var(U),\var(V)\subseteq\dom(\sigma)$ an
d $M$ ground. Let $W, W'$ be subterms of terms in $\ran(\sigma)$ such that for every occurrence $q_{\x}$
of $\x$ in $W$ (or $W'$) there is an occurrence of an encryption $q_{\enc}$ in $W$ (or $W'$ respectively)
with $q_{\enc}<q_{\x}$.
Then
\begin{enumerate}
 \item\label{corol_condsubst} $U\sigma[\subst{\x}{M}] = V\sigma[\subst{\x}{M}]$ implies $U\sigma = V\sigma$;
 \item\label{corol_condmixt} $U\sigma[\subst{\x}{M}] = W[\subst{\x}{M}]$ implies $U\sigma = W$;
 \item\label{corol_condsub} $W[\subst{\x}{M}] = W'[\subst{\x}{M}]$ implies $W = W'$.
\end{enumerate}
\end{cor}
\proof
We prove below that in $U\sigma$ and in $W$  for each occurrence $q_{\x}$ of $\x$ there is an
encryption $q'_{\enc}$ (in $y\sigma$ for some $y\in\var(U)$, and in $W$ respectively) with $q'_{\enc}\pdot
1\le q_{\x}$ such that $q'_{\enc}$ is an agent encryption w.r.t.~$\nb\!\setminus\!\!\set{\x}$. Then,  by
analogy, the same thing holds for $V\sigma$ and $W'$. Since by condition (2) of extended well-formed
frames an agent encryption w.r.t.~$\nb$ is a probabilistic encryption, it follows that each pair $(U\sigma,
V\sigma)$, $(U\sigma, W)$ and $(W,W')$ satisfies the conditions of Lemma~\ref{lemma_synt}. Then the result
follows directly.



Consider an occurrence $q_{\x}$ of $\x$ in $U\sigma$. Since $U$ is public, there is a
variable $y\in\var(U)\subseteq\dom(\sigma)$ and an occurrence $p_y$ of it in $U$ such that $p_y\le q_{\x}$.
From the definition of extended well-formed frames we know that there is an encryption $q'_{\enc}$ in
$y\sigma$ with $q'_{\enc}\pdot 1\le q_{\x}$ which is an agent encryption {w.r.t.}
$\nb\!\setminus\!\!\set{\x}$. 
Hence $q'_{\enc}$ satisfies the conditions of Lemma~\ref{lemma_synt}.

In $W$ for each occurrence $q_{\x}$ of $\x$ there is an occurrence $q_{\enc}$ of an encryption above
$q_{\x}$. Then we can consider the lowest occurrence $q'_{\enc}$ of an encryption above $q_{\x}$ in $W$. By
the definition of extended well-formed frames, the lowest encryption above $q_{\x}$is an agent encryption
and is plain-text above $q_{\x}$. Hence $q'_{\enc}$ satisfies the conditions of Lemma~\ref{lemma_synt}.\qed

Lemma~\ref{lemma_syntactic} can now be easily deduced  since it is the analogous
statement of Point~\ref{corol_condsubst} of Corollary \ref{corol_synt} for well-formed frames (which are
extended well-formed frames as we have seen in Lemma~\ref{lem:well-ext}).

\newcommand{\lemmasimulaterev}{
Let $\varphi=\nu\nb.\sigma$ be an extended well-formed frame w.r.t.~$\x\in\nb$ such that $\varphi\nvdash \x$.
Let  $U$ be a term with $\var(U)\subseteq\dom(\varphi)$ and $M$ be
a closed term in normal form such that $U$ and $M$ are public w.r.t.~$\varphi$. If
$U\sigma[\subst{\x}{M}]\rightarrow V$, for some term $V$, then there exists an extended
well-formed frame $\varphi'=\nu\nb.\sigma'$ w.r.t.~$\x$
\begin{itemize}
\item extending $\varphi$, that is $x\sigma'=x\sigma$ for all $
  x\in\dom(\sigma)$,
\item preserving deducible terms: $\varphi\vdash W$ if and only if
  $\varphi'\vdash W$,
\item and such that  $V=V'\sigma'[\subst{\x}{M}]$ and $U\sigma\rightarrow V'\sigma'$ for some $V'$
public w.r.t.~$\varphi'$.
\end{itemize}
}

The following lemma is the generalization of Lemma~\ref{lemma_simulate} for extended well-formed frames.
\begin{lem}\label{lemma_simulate_rev}
\lemmasimulaterev
\end{lem}

We give here only a proof sketch, the detailed proof can be found in
Appendix~\ref{proof_lemma_simulate_rev}.

\proof[Proof sketch]
Let $U,V,M$ be terms with $U$ and $M$ public w.r.t.~$\varphi$, $M$ being closed and in normal form such that
$U\sigma[\subst{\x}{M}]\rightarrow V$, as in the statement of the lemma. Let $L\rightarrow
R\in\mathcal{R}_E$ be the rule that was applied in the above reduction and let $p$ be the position at which
it was applied, \textit{i.e.}~$U\sigma[\subst{\x}{M}]|_p=L\theta$.
Since $M$ is in normal form, $p\in\pos(U\sigma)$.

By a case analysis of the rewrite rules in $\mathcal{R}_E$ one can prove that there is a substitution
$\theta_0$ such that $U\sigma|_p=L\theta_0$. It follows that $U\sigma$ is reducible.
Since all terms in an extended-well formed frame, thus in $\varphi$, are in normal form, we have that
$p\in\posnv(U)$. Then, for $T=U|_p$, $T\sigma[\subst{\x}{M}]=L\theta$ and $T\sigma=L\theta_0$.

For our equational theory $E$, $R$ is either a constant (\emph{i.e.}~$\ok$) or a
variable.  If $R$ is a constant then we take $V'=U[R]_p$ and
$\sigma'=\sigma$. If $R$ is a variable, say $z_0$, then consider the position $q$ of $z_0$ in $L$. This
position $q$ is also in $L\theta_0$, that is in $T\sigma$.
Hence the two following possibilities may occur:
\begin{enumerate}
\item If $q\in\posnv(T)$, that is there is no $y\in\dom(\sigma)$ above
$z_0$, then we consider $V'=U[T|_q]_p$ and $\sigma'=\sigma$. 

\item If $q\notin\posnv(T)$, that is there is some
$y\in\dom(\sigma)$ above $z_0$, then we consider $V'=U[y']_p$ and
$\sigma'=\sigma\cup\{R\theta_0/y'\}$, where $y'$ is a new variable (i.e.\ $y'\notin\dom(\sigma)$).
\end{enumerate}
A simple analysis of these three cases shows that $\sigma'$ and $V'$ satisfy that the conditions of the
lemma.\qed

\section{Active case}\label{sec:active}

In the active case, we provide sufficient conditions for syntactic and
strong secrecy to be also equivalent. In particular, we require that
no test is performed directly on the secret. 
We establish our equivalence result in the applied pi calculus framework,
introduced by Mart\`in Abadi and C\'edric Fournet.
We do not make any restriction on the use of the replication symbol,
which means that protocols with an unbounded number of sessions as
well as protocols with a bounded number of sessions 
can be considered.

\comment{
To simplify the analysis of the active case, we
restrict our attention to pairing and symmetric encryption: the
alphabet $\Sigma$ is now reduced to $\Sigma = \{\enc, \dec,
 \langle\rangle, \pi_1, \pi_2\}$ and $E$ is limited to the first three equations.
}

\subsection{Modeling protocols within the applied pi calculus}



\comment{All other definitions are as in Abadi-Fournet's applied pi
calculus.} The applied pi calculus~\cite{AbadiFournet01} is a
process algebra well-suited for modeling cryptographic protocols,
generalizing the spi-calculus~\cite{abadi97calculus}. We shortly
describe its syntax and semantics. This part is mostly borrowed
from~\cite{AbadiFournet01}.

\smallskip
\emph{Processes}, also called plain processes, are defined by the grammar:
\[
\begin{array}{l}
\hspace*{-0.25cm}P, Q\  :=\   \text{processes}
\\
\begin{array}{l@{\hspace{0.25cm}}l@{\hspace{1cm}}l@{\hspace{0.25cm}}l}
\mathbf{0} & \text{null process}&
\nu n.P & \text{name restriction}\\
P\mid Q & \text{parallel composition}
& u(z).P & \text{message input}\\
!P & \text{replication}
& \out{u}{M}.P & \text{message output}\\
\mathit{if}\ T=T'\ \mathit{then}\ P\ \mathit{else}\ Q\ & \text{conditional}\\
\end{array}\end{array}\]
where $n$ is a name, $M$, $T$, $T'$ are terms, and $u$ is a name or a variable. The null process~$\mathbf{0}$ does
nothing. Parallel composition executes the two processes concurrently. Replication $!P$ creates unboundedly
many instances of $P$. Name restriction $\nu n.P$ builds a new, private name $n$, called \emph{channel name},
binds it in $P$ and then
executes $P$. The conditional $\mathit{if}\ T=T'\ \mathit{then}\ P\ \mathit{else}\ Q$ behaves like $P$
or $Q$ depending on the result of the test $T=T'$. If $Q$ is the null process then we use the notation
$[T=T'].P$ instead.  Finally, the process $u(z).P$ inputs a message and executes $P$ binding the variable $z$
to the received message, while the process $\out{u}{M}.P$ outputs the message $M$ and then behaves like $P$.
We may omit $P$ if it is $\mathbf{0}$.
In what follows, we restrict our attention to the case where $u$ is a
name since it is usually sufficient to model cryptographic protocols.%
\footnote{Note that we do not change the calculus. In particular,
  there is no restriction on the use of channels for
  adversaries/observers that are used in the definition of
  observational equivalence.}

\smallskip
\emph{Extended processes} are defined by the grammar:
\[\begin{array}{l}
\hspace*{-0.4cm}A, B\  :=\   \text{extended processes}\\
\begin{array}{l@{\hspace{0.4cm}}l@{\qquad\qquad}l@{\hspace{0.4cm}}l}
P & \text{plain process}&
\nu n.A & \text{name restriction}\\
A\mid B & \text{parallel composition}& \nu x.A & \text{variable restriction}\\
\set{\subst{x}{M}} & \text{active substitution}\\
\end{array}\end{array}\]
\emph{Active substitutions} are just cycle-free substitutions. They generalise the $\mathit{let}$ binding,
in the sense that $\nu x.(\set{\subst{x}{M}}|P)$ corresponds to $\mathit{let}\ x=M\ \mathit{in}\ P$ standard
construction, while unrestricted, $\set{\subst{x}{M}}$ behaves like a permanent
knowledge, permitting to refer globally to $M$ by means of $x$. Substitutions
$\set{\subst{x_1}{M_1},\dots,\subst{x_l}{M_l}}$ with $l\ge 0$ are identified with extended processes
$\set{\subst{x_1}{M_1}}|\ldots|\set{\subst{x_l}{M_l}}$. In particular, the empty substitution is identified
with the null process.

We denote by $\fv(A)$, $\bv(A)$, $\fn(A)$, and $\bn(A)$ the sets of  free and bound variables and free and
bound names of $A$, respectively, defined inductively as usual
and using
$\fv(\set{\subst{x}{M}}) = \fv(M) \cup \set{x}$ and $\fn(\set{\subst{x}{M}}) = \fn(M)$ for active
substitutions. An extended process is \emph{closed} if it has no free variables except those in the domain
of active substitutions.

Extended processes built up from the null process and active substitutions (using the given
constructions, that is, parallel composition, restriction and  active
substitutions) are called \emph{frames}\footnote{We see
later in this section why we use the same name as for the notion defined in
Section~\ref{sec:passive}.}. To every extended process $A$ we
associate the  frame $\varphi(A)$ obtained by replacing all embedded
plain processes with $\mathbf{0}$. For example, if $A=\nu y,k,r.\set{\subst{x}{\enc(m,k,r)},\subst{y}{a}}\mid \out{c}{y}$
then $\varphi(A)=\nu
y,k,r.\set{\subst{x}{\enc(m,k,r)},\subst{y}{a}}$. 
Note that $\varphi(A)\equiv\nu k,r.\set{\subst{x}{\enc(m,k,r)}}$.

An \emph{evaluation context} is an extended process with a hole not
under a replication, a conditional, an input or an output.

\emph{Structural equivalence} ($\equiv$) is the smallest equivalence relation on extended processes that is
closed by $\alpha$-conversion of names and variables, by application of evaluation contexts and such that
the standard structural rules for the null process, parallel composition and restriction (such as
associativity and commutativity of $|$, commutativity and binding-operator-like behaviour of $\nu$) together
with the following ones hold.

\[\begin{array}{r@{\hspace*{0.15cm}}l@{\hspace*{1cm}}r}
\nu x.\set{\subst{x}{M}}\equiv & \mathbf{0} & \mbox{\regle{ALIAS}}\medskip\\
\set{\subst{x}{M}}\,|\, A \equiv & \set{\subst{x}{M}}\, |\, A\set{\subst{x}{M}} & \mbox{\regle{SUBST}}\medskip\\
\set{\subst{x}{M}}\equiv & \set{\subst{x}{N}} \mbox{\quad if $M=_E N$} & \mbox{\regle{REWRITE}}
  \end{array}\]

If $\nb$ represents the (possibly empty) set $\set{n_1,\ldots,n_k}$, we abbreviate by $\nu\nb$ the sequence
$\nu n_1.\nu n_2\dots\nu n_k$. Every closed extended process $A$ can be brought to the form
$\nu\nb.\set{\subst{x_1}{M_1}}|\ldots|\set{\subst{x_l}{M_l}}|P$ by using structural equivalence, where
 $P$ is a plain closed process, $l\ge 0$ and $\nb\subseteq\cup_{i}\fn(M_i)$.  Hence the two definitions
of frames are equivalent up to structural equivalence on closed extended processes. To see this we apply
rule \regle{SUBST} until all terms are ground (this is assured by the fact that the considered extended
processes are closed and the active substitutions are cycle-free). Also, another consequence is that if
$A\equiv B$ then $\varphi(A) \equiv \varphi(B)$.

Two semantics  can be considered for this calculus, defined by
structural equivalence and by \emph{internal reduction} and
\emph{labeled reduction}, respectively. These semantics lead to
\emph{observational equivalence} (which is standard and not recalled
here) and \emph{labeled bisimilarity} relations. The two
bisimilarity relations are equal~\cite{AbadiFournet01}. We use here
the latter since it relies on static equivalence and it allows to
take implicitly into account the adversary, hence having the
advantage of not using quantification over contexts.

\smallskip
\emph{Internal reduction} is the smallest relation on extended processes which is closed by structural
equivalence and
application of evaluation contexts, and such that:
\[\begin{array}{r@{\hspace*{0.2cm}}l@{\hspace*{1.2cm}}r}
\out{c}{x}.P \mid c(x).Q\ \rightarrow & P \mid Q & \mbox{\pregle{COMM}}\medskip\\
\mathit{if}\ T=T'\ \mathit{then}\ P\ \mathit{else}\ Q\  \rightarrow
& P & \mbox{\pregle{THEN}}\\
\multicolumn{3}{l}{\mbox{\quad\phantom{}for any ground terms $T$ and $T'$ such that  $T =_E T'$}}\medskip\\
\mathit{if}\ T=T'\ \mathit{then}\ P\ \mathit{else}\ Q\  \rightarrow & Q  &
\mbox{\pregle{ELSE}}\\
\multicolumn{3}{l}{\mbox{\quad\phantom{}for any ground terms $T$ and $T'$ such that  $T \neq_E T'$}} 
\end{array}\]

\smallskip

On the other hand, \emph{labeled reduction} is defined by the following rules:
\[\begin{array}{l@{\hspace{0.2cm}}l@{\quad}l@{\hspace{0.3cm}}l}
c(x).P
\xrightarrow{c(M)} P\set{\subst{x}{M}} & \mbox{\pregle{IN}}&
\out{c}{u}.P\stackrel{\out{c}{u}}{\longrightarrow}P & \mbox{\pregle{OUT-ATOM}}\\

\infer[\footnotesize u\neq c]{\nu u.A
\xrightarrow{\nu u.\out{c}{u}}A'}{A
\xrightarrow{\out{c}{u}} A'} & \raisebox{4.3mm}{\mbox{\pregle{OPEN-ATOM}}} &
\infer[\parbox{5em}{\footnotesize $u$ does not\\ occur in $\alpha$}]
{\nu u.A\stackrel{\alpha}{\longrightarrow}\nu u.A'}{A\stackrel{\alpha}{\longrightarrow}A'}
&
\raisebox{3.8mm}{\text{\pregle{SCOPE}}}\\

\infer[\mbox{
(*)}]{A|B\stackrel{\alpha}{\longrightarrow}A'|B}{A\stackrel{\alpha}{\longrightarrow}A'}
 & \raisebox{3mm}{\pregle{PAR}}&
\infer{A\stackrel{\alpha}{\longrightarrow}A'}{A\equiv B \quad
B\stackrel{\alpha}{\longrightarrow}B' \quad
B'\equiv A'} & \raisebox{3mm}{\pregle{STRUCT}}\\
\end{array}\]
where $c$ is a name and $u$ is a metavariable that ranges over names and variables, and the condition (*) of
the rule \regle{PAR} is $\bv(\alpha)\cap\fv(B)=\bn(\alpha)\cap\fn(B)=\emptyset$.

\begin{definition}\label{def_lab_bisim}
\emph{Labeled bisimilarity} ($\approx_l$) is the largest symmetric relation $\mathcal{R}$ on closed extended
processes such that $A\,\mathcal{R}\, B$ implies:
\begin{enumerate}
 \item $\varphi(A)\approx \varphi(B)$;
 \item if $A\rightarrow A'$ then $B\rightarrow^* B'$ and
$A'\,\mathcal{R}\,B'$, for some~$B'$;
 \item if $A\lab A'$ and $\fv(\alpha)\subseteq\dom(\varphi(A))$ and $\bn(\alpha)\cap\fn(B)=\emptyset$
then $B\rightarrow^\ast\lab\rightarrow^\ast B'$ and $A'\,\mathcal{R}\, B'$, for some $B'$.
\end{enumerate}
\end{definition}

We denote $A \Rightarrow B$ if $A \rightarrow B$ or $A\lab B$.
\begin{definition}
A frame $\varphi$ is \emph{valid} {w.r.t.} a process $P$ if there is $A$ such that $P\Rightarrow^* A$ and
$\varphi\equiv\varphi(A)$.
\end{definition}

\begin{definition}
Let $P$ be a closed plain process without variables as channels and $\x$ a bound name of $P$, but
not a channel name. We say that  $\x$ is \emph{\syntactically secret} in $P$ if, for every valid frame
$\varphi$
{w.r.t.}~$P$, $\x$ is not deducible from $\varphi$. We say that $\x$ is \emph{strongly secret} if for
any closed terms $M,M'$ such that $\bn(P)\cap(\fn(M)\cup\fn(M'))=\emptyset$, $P[\subst{\x}{M}]\approx_l
P[\subst{\x}{M'}]$,
where $P[\subst{s}{M}]$ represents the instantiation of the name $s$ with $M$ in $P$
except (of course) in the name restriction constructions.
\end{definition}

Let $\mathcal{M}_o(P)$ be the set of \emph{outputs} of $P$, that is the set of terms $m$ such that
$\out{c}{m}$ is a message output construct  for some channel name $c$ in $P$, and let $\mathcal{M}_t(P)$ be
the set of \emph{operands of tests} of $P$, where a \emph{test} is a
couple $T=T'$ occurring in a conditional and its  \emph{operands} are $T$ and $T'$.
Let $\mathcal{M}(P)=\mathcal{M}_o(P)\cup\mathcal{M}_t(P)$ be the
set of \emph{messages} of~$P$.
Examples are provided  at the end of this section.

\newcommand{\lemmaframe}{
Let $P$ be a closed plain process, and $A$ be a closed extended
process such that $P\Rightarrow^* A$. There are $l\ge 0$, an
extended process $B=\nu\nb.\sigma_l|P_B$, where $P_B$ is some plain
process, and $\theta$ a substitution public {w.r.t.} $\nb$ such
that: $A\equiv B$, $\nb\subseteq\bn(P)$, for every operand of
a test or an output $M$ of $P_B$ there is a message $M_0$ in $P$
(an operand of a test or an output respectively), such that
$M=M_0\theta\sigma_l$, and, $\sigma_{i}=\sigma_{i-1}\cup\lbrace
\subst{y_i}{M_{i}\theta_i\sigma_{i-1}}\rbrace$ is a ground substitution,
for all $1\le i\le l$, where $M_i$ is an output in $P$, $\theta_i$ is a substitution
public {w.r.t.} $\nb$ and $\sigma_0$ is the empty substitution.
}

The following lemma intuitively states that any message contained in
a valid frame is an output instantiated by messages deduced from
previous sent messages.
\begin{lem}\label{lemma_frame}
\lemmaframe
\end{lem}

The proof is done by induction on the number of reductions in $P\Rightarrow^* A$.
A detailed proof can be found in Appendix~\ref{proof_lemma_frame}.
Intuitively, $B$ is obtained by applying the \regle{SUBST} rule (from left to right) as much as
possible until there are no variables left in the plain process. Note that $B$ is unique up to the
structural rules different from \regle{ALIAS}, \regle{SUBST} and \regle{REWRITE}. We say that $\varphi(B)$ is
the \emph{standard frame} {w.r.t.} $A$.


\comment{The lemma implies that no "let" structures are used when writing a process in its standard form.}


As a running example we consider the Yahalom protocol:
\[\begin{array}[c]{r@{\hspace*{0.2cm}}l}
A\Rightarrow B: & A,N_a \\
B\Rightarrow S: & B,\penc{A,N_a,N_b}{K_{bs}} \\
S\Rightarrow A: & \penc{B,K_{ab},N_a,N_b}{K_{as}},\penc{A,K_{ab}}{K_{bs}} \\
A\Rightarrow B: & \penc{A,K_{ab}}{K_{bs}}\comment{,\penc{N_b}{K_{ab}}} \\
\end{array}
\]
In this protocol, two participants $A$ and $B$ wish to establish a shared key $K_{ab}$. The key is created
by a trusted server $S$ which shares the secret keys $K_{as}$ and $K_{bs}$ with $A$ and $B$ respectively.
The protocol is modeled by the following process:
$$P_Y = \nu k_{as}, k_{bs}.\,(!P_A)\mid(!P_B)\mid(! \nu k.P_S(k))\mid\nu k_{ab}.P_S(k_{ab})$$
with
 \[\begin{array}[c]{l}
P_A=\nu n_a.\out{c}{a,n_a}.c(z_a).
   [ b=U_b ]. [ n_a=U_{n_a} ].\out{c}{\pi_2(z_a)}.\mathbf{0}\\
P_B=c(z_b).\nu
n_b,r_b.\out{c}{b,\enc(\pair{\pi_1(z_b)}{\pair{\pi_2(z_b)}{n_b}},k_{bs},r_b)}.c(z'_b).[a=\pi_1(\dec(z'_b,k_{
bs}))].\mathbf{0}\\
P_S(x)=c(z_s).[a=V_a].[b=\pi_1(z_s)].\nu r_s,r'_s.\\
\qquad\qquad\qquad\qquad\qquad\qquad\out{c}{\pair{\enc(\pair{\pi_1(z_s)}{\pair{x}{V_n}},k_{as},r_s)}{
\enc(\pair{V_a}{x}, k_{bs},r'_s)} } .\mathbf { 0 }\\
\end{array}\]
where \quad$\begin{array}[t]{l@{\qquad\qquad}l}
U_b=\pi_1(\dec(\pi_1(z_a),k_{as}))&
U_{n_a}=\pi_1(\pi_2(\pi_2(\dec(\pi_1(z_a),k_{as}))))\\
V_a=\pi_1(\dec(\pi_2(z_s),k_{bs}))&
V_n=\pi_2(\dec(\pi_2(z_s),k_{bs})).\\
\end{array}$\\[0.1cm]

Note that for simplicity and
concision, we only consider two honest agents. However, we could
extend the process to the case where A and B are also willing to
interact with a corrupted identity C and establish a similar result.

For this protocol the set of outputs and operands of tests are
respectively:
\[\begin{array}{rl}
\mathcal{M}_o(P_Y)=& \{\pair{a}{n_a},\pi_2(z_a),
\pair{b}{\enc(\pair{\pi_1(z_b)}{\pair{\pi_2(z_b)}{n_b}},k_{bs},r_b)},\\
&
\hspace{0.2cm}\pair{\enc(\pair{\pi_1(z_s)}{\pair{x}{V_n}},k_{as},r_s)}{\enc(\pair{V_a}{x},k_{bs},r'_s)}\} \mbox{ and
}
\\

\mathcal{M}_t(P_Y)=& \set{b,U_b,n_a,U_{n_a},a,\pi_1(\dec(z'_b,k_{bs})),V_a,b,\pi_1(z_s)}.
\end{array}\]

\subsection{Our hypotheses}\label{sec-hyp}

In what follows, we assume $\x$ to be the desired secret.
As in the passive case, destructors above the secret must be forbidden.
We also restrict ourself to processes
with ground terms in key position. 
Indeed, consider the process \[P_1=\nu \x,k,r,r'.\bigl(\out{c}{\enc(\x,k,r)}\,|\,
c(z).\out{c}{\enc(a,\dec(z,k),
 r')}\bigr).\]
The name $\x$ in $P_1$ 
is \syntactically secret but not strongly secret. Indeed,
\[\begin{array}{@{}rl}
P_1& \equiv\   \nu \x,k,r,r'.\bigl(\nu z.\bigl(\{\subst{z}{\enc(\x, k, r)}\}
\mid \out{c}{z}\mid  c(z).\out{c}{\enc(a, \dec(z,k), r')}\bigr)\bigr)\\
& {\rightarrow}\  \nu \x,k,r,r'.\bigl(\{\subst{z}{\enc(\x, k, r)}\}\mid\out{c}{\enc(a, \x,
r')}\bigr)
\qquad\qquad\mbox{ (\regle{COMM} rule)}\\
& \equiv\  \nu \x,k,r,r'.\bigl(\nu z'.\bigl(\{\subst{z}{\enc(\x,k,r)},
  \subst{z'}{\enc(a,\x,r')}\}
\mid \out{c}{z'}\bigl)\bigr)\\
& \xrightarrow{\nu z'.\out{c}{z'}}\  P'_1 = \nu
\x,k,r,r'.\{\subst{z}{\enc(\x,k,r)},
  \subst{z'}{\enc(a,\x,r')}\}
\end{array}
\]
and $P_1'$ does not preserve the strong secrecy of $\x$ (see the frame $\psi_2$ of Section~\ref{sec:implic}).


Without loss of generality with respect to cryptographic protocols, 
 we assume that terms occurring in processes  are in normal form and that no destructor appears above
constructors. Indeed, terms like $\pi_1(\encg(M,K,R))$ are usually not used to specify protocols. We also
assume that tests do not contain constructors. Indeed a test $[\langle T_1,T_2\rangle = T']$ can be rewritten
as $[T_1 = T'_1].[T_2=T'_2]$ if $T'= \langle T'_1,T'_2\rangle$, and $[T_1 = \pi_1(T')].[T_2 = \pi_2(T')]$ if
$T'$ does not contain constructors, and will never hold otherwise. Similar rewriting applies for encryption,
except for the test $[\encg(T_1,T_2,T_3)= T']$ if $T'$ does not contain constructors. It can be rewritten in
$[\decg(T',T_2) = T_1]$ but this is not equivalent. However since the randomness of encryption is not known
to the agents, explicit tests on the randomness should not occur in general.

This leads us to consider the following class of processes.
\begin{definition}\label{def_well-formed}
A process $P$ is \emph{well-formed} w.r.t.~a name $\x$ if it is closed, channels are names different from $\x$ and:
\begin{enumerate}
\item \label{cond_retr_check}
the symbol $\retr$ does not occur in $\mathcal{M}(P)$, the symbol
$\scheck$ does not occur in $\mathcal{M}(P)$ except in head of a test,
that is, the check symbol can only appear in tests of the form 
$[\scheck(M,N,K)=\ok]$ where $\scheck$ does not appear
in $M,N,K$;
 \item \label{cond_probenc}
any encryption in some term of $\mathcal{M}(P)$ is a probabilistic agent encryption w.r.t.~$\mathcal{M}(P)$
and $\bn(P)\!\setminus\!\set{\x}$ respectively;
\comment{
any occurrence of $\encg(M,K,R)$ in some term $T\in\mathcal{M}(P)$ is an agent encryption {w.r.t.}~$\bn(P)$,
and for any term $T'\in \mathcal{M}(P)$ and position $p$ such that $T'|_p=T$  there is a position $q$ such
that $q.3=p$ and $T'|_q= \encg(M,K,R)$;}
 \item \label{cond_keys}
for any subterm term $\encg(M,K,R)$, $\decg(M,K)$ or $\sign(M,K)$ occurring in $\mathcal{M}(P)$, $K$ is a closed
term;
 \item \label{cond_no_dest} in $\mathcal{M}(P)$ there are no destructors, nor $\pub$ or $\priv$ function
symbols above constructors, nor above $\x$;
 \item \label{cond_test_message}
for any test,
\begin{itemize}
\item either each operand of a test $T\in\mathcal{M}_t$ is a name, a constant or
has the form $\pi^{1}(\dec_1(\dots\pi^{l}(\dec_l(\pi^{l+1}(z),K_l))\dots,K_1))$,
with $l\ge 0$, where $\dec_i\in\set{\dec,\deca}$, $\pi^{i}$ are words on $\{\pi_1,\pi_2\}$ and $z$ is
a variable,
\item or the  test is $[\scheck(M,N,K)=\ok]$ with $K$ being a closed term and $M$ and $N$
is of the previously described form.
\end{itemize}
\end{enumerate}
\end{definition}

Conditionals should not test on $\x$. For example, consider the following process:
\[P_2= \nu \x,k,r.\bigl(\out{c}{\enc(\x,k,r)}\mid c(z).[\dec(z,k) = a].\out{c}{\ok}\bigr)\]
where $a$ is a non restricted name. The name $\x$ in $P_2$ is \syntactically secret but not strongly secret.
Indeed, $P_2 \rightarrow \nu \x,k,r.(\{\subst{z}{\enc(\x,k,r)}\}\,|\,[\x=a].\out{c}{\ok})$ and the process
$P_2[\vsubst{\x}{a}]$ reduces further, while $P_2[\vsubst{\x}{b}]$ does not.


That is why we have to prevent hidden tests on $\x$. Such tests may occur nested in equality tests. For
example, let
\[\begin{array}{l@{}l}
P_3 = & \nu {\x,k,r, r_1,r_2}.\bigl(\out{c}{\enc(\x,k,r)}
\mid \out{c}{\enc(\enc(a,k',r_2),k,r_1)} \\
& \qquad\qquad\qquad\qquad\qquad
\mid c(z).[\dec(\dec(z,k),k') = a].\out{c}{\ok}\bigl)\quad \rightarrow  \\
P'_3\, = & \ \nu {\x,k,r, r_1,r_2}.\bigl(\set{\subst{z}{\enc(\x,k,r)}}
\mid  \out{c}{\enc(\enc(a,k',r_2),k,r_1)} \mid  [\dec(\x,k') = a].\out{c}{\ok}\bigr)
\end{array}\]
Then $P_3[\subst{\x}{\enc(a,k',r')}]$ is not equivalent to $P_3[\subst{\x}{n}]$, since the process
$P'_3[\subst{\x}{\enc(a,k',r')}]$ emits the message $\ok$ while $P'_3[\subst{\x}{n}]$ does not. This relies
on the fact that the decryption $\dec(z,k)$ allows access to $\x$ in the test.



For the remaining of the section we assume that $\y$ and $\z$ are new fixed variables.
To prevent hidden tests on the secret, we compute an over-approximation of the ciphertexts that may contain
the secret, by marking with $\y$ all positions under which the secret may appear
in clear. 

We first introduce a function $f_{ep}$ that extracts the lowest encryption over $\x$ and ``cleans up'' the
pairing function above $\x$.
Formally, we define the partial function
$$\begin{array}{rcl}f_{ep}\colon\mathcal{T}\times
\mathbb{N}_+^*&\hookrightarrow&\mathcal{T}\times \mathbb{N}_+^*
  \end{array}
$$
$f_{ep}(U,p) = (V,q)$ where $V$ and $q$ are defined as follows:
$q\leq p$ is the position (if it exists) of the lowest encryption on the path $p$ in
 $U$. If $q$ does not exist or if $p$ is not a maximal position in
 $U$, then $f_{ep}(U,p) = \perp$.
Otherwise, $V$ is obtained from
$U|_q$ by replacing all arguments of pairs that are not on the path
$p$ with new variables.
More precisely, let $V'= U|_q$. The subterm $V'$ must be of the form
$\encg(M_1,M_2,M_3)$ and $q=i\pdot q'$. If $i\neq 1$, then
$f_{ep}(U,p) = \perp$.
Otherwise, $V$ is defined by $V = \encg(M_1',M_2 ,M_3)$ with 
$M_1' = \simplify(M_1,q')$ where $\simplify$ is recursively
defined by:
\[\begin{array}{l}
\simplify(N,\epsilon) =  N\\
\simplify(\langle N_1,N_2\rangle,1\pdot r) = \langle\simplify(N_1,r), x_{2\pdot r}\rangle\\
\simplify(\langle N_1,N_2\rangle,2\pdot r) =\! \langle x_{1\pdot r}, \simplify(N_2,r)\rangle\\
\simplify(\sign(M,K),1\pdot r) = \sign(\simplify(M),x_{2\pdot r})\\
\simplify(f(N_1,\ldots,N_k),r) = f(N_1,\ldots,N_k)\quad \mbox{if $f$
  is a destructor}\\
\end{array}\]
and is undefined in all other cases.
For example,
\psset{levelsep=20pt,nodesep=1pt}
\[
f_{ep}(
\pstree{\Tr{\enc}}{
  \pstree{\Tr{\enc}}{
    \pstree{\Tr{\pair{}{}}}{
      \pstree{\Tr{\pair{}{}}}{
    \TR{a}
    \TR{b}
      }
      \TR{c}
    }
    \TR{k_2}
    \TR{r_2}
  }
  \TR{k_1}
  \TR{r_1}
}
, 1\pdot 1\pdot 2) =
(\pstree{\Tr{\enc}}{
    \pstree{\Tr{\pair{}{}}}{
      \TR{z_{1\pdot 2}}
      \TR{c}
    }
    \TR{k_2}
    \TR{r_2}
  }
,1)
\]

The function $f_e$ is the composition of the first projection with $f_{ep}$. With the function $f_e$, we can
extract from the outputs of a
protocol $P$ the set of ciphertexts where $\x$ appears explicitly
below the encryption.
$$\mathcal{E}_0(P)=\set{f_e(M[\y]_p,p)\mid M\in\mathcal{M}_o(P)\
\wedge\ M|_p=\x}.$$
For example,
$\mathcal{E}_0(P_Y)=\set{\enc(\pair{z_{1\pdot 1}}{\pair{\y}{z_{2}}},k_{as},r_s),
  \enc(\pair{z_{1}}{\y},k_{bs},r'_s)}$,
where $P_Y$ is the process corresponding to the Yahalom protocol
  defined in previous section. 

However $\x$ may appear in other ciphertexts sent later on during the
execution of the protocol after decryptions and encryptions. Thus we also
extract from outputs the destructor parts (which may open encryptions). Namely, we define the partial
function
\[f_{dp}\colon\mathcal{T}\times\mathbb{N}_+^*\hookrightarrow\mathcal{T}\times\mathbb{N}_+^*
\]
$f_{dp}(U,p)= (V,q) $ where $V$ and $q$ are defined as follows: $q\leq p$ is the occurrence of the highest
destructor different from $\scheck$ above $p$ (if it exists). Let $r\leq p$ be the occurrence of the lowest decryption
above $p$ (if
it exists). We have $U|_r = \decg(U_1,U_2)$. Then $U_1$ is replaced by
the variable $\z$ that is $V=(U[\decg(\z,U_2)]_{r})|_q$. If $q$ or $r$ do not exist then $f_{dp}(U,p)= \perp
$.

For example,
$
f_{dp}(\enc(\pi_1(\dec(\pi_2(y),k_1)),k_2,r_2),1\pdot 1\pdot 1\pdot 1) = (\pi_1(\dec(\z,k_1)),1).
$

The function $f_d$ is the composition of the first projection with $f_{dp}$. By applying the function $f_d$
to messages of a well-formed process $P$ we always obtain either terms $D$ of the form\footnote{in this context
we simply write $D(T)$ instead of $D[\subst{\z}{T}]$} $D=D_1(\dots D_n)$ where
$D_i(\z)=\pi^i(\decg(\z,K_i))$ with $1 \leq i \leq n$, $K_i$ are ground terms and $\pi^i$ is a (possibly empty)
sequence of projections $\pi_{j_1}(\pi_{j_2}(\dots(\pi_{j_l})\dots))$,
or terms $\scheck(M,D,K)$ where $D$ is of the previously defined form.


With the function $f_{d}$, we can extract from the outputs of a protocol $P$ the meaningful destructor part.
\[\mathcal{D}_o(P)=\set{f_d(M,p)\mid M\in\mathcal{M}_o(P)\ \wedge\ p\in\posv(M)}.
\]

Remember that $\posv(M)$ is the set of variable positions.

For example,
$\mathcal{D}_o(P_Y)=\set{\pi_2(\dec(\z,k_{bs})),\pi_1(\dec(\z,k_{bs}))}$.

We are now ready to mark (with $\y$) all the positions where the secret might be transmitted (thus tested).
We define inductively the sets $\mathcal{E}_i(P)$ as follows. For each element $E$ of $\mathcal{E}_i$
we can show that there is an unique term in normal form denoted by  $\bb{E}$ such that
$\var(\bb{E})=\set{\z}$ and $\nf{\bb{E}(E)}\!=\!\y$.
That is, intuitively, $\bb{E}$ opens $E$ until $\y$.
For example, let $E_1\! =\!
\enc(\pair{z_1}{\pair{\y}{z_2}},k_{as},r_s)$, then $\bb{E_1} = \pi_1(\pi_2(\dec(\z,k_{as})))$.
We define
\[\begin{array}{rcl}
 \bb{\mathcal{E}_i}(P)  &=& \{U\mid \exists E\in\mathcal{E}_i(P),U\le_{st}\bb{E}\ \mbox{ and }
 \exists q\in\pos(U),\head{U|_q}=\decg\},\\
\mathcal{E}_{i+1}(P) &=& \{M'[\y]_q\mid \exists M\in\mathcal{M}_o(P),p\in\posv(M)
 \mbox{ s.t. }f_{ep}(M,p)=(M',p'), \\&&\qquad\quad f_{dp}(M',p'')=(D,q),
 p=p'\pdot p'', D=D_1(\dots D_n), \mbox{ and } D_1\in\bb{\mathcal{E}}_i(P) \}.
\end{array}
\]
For example,\\
$\begin{array}{l@{\hspace{0.2em}}l@{\hspace{0.2em}}l}
\bb{\mathcal{E}_0}(P_Y)&=&\{\pi_1(\pi_2(\dec(\z,k_{as}))),\pi_2(\dec(\z,k_{as})),
\dec(\z,k_{as}),\pi_2(\dec(\z,k_{bs})),\dec(\z,k_{bs})\}\\
\mathcal{E}_1(P_Y)&=&\set{\enc(\pair{z_{1\pdot 2}}{\pair{z_{1}}{\y}},k_{as},r_s)}\\
\bb{\mathcal{E}_1}(P_Y)&=&\{\pi_2(\pi_2(\dec(\z,k_{as}))),
\pi_2(\dec(\z,k_{as})),\dec(\z,k_{as})\}\end{array}$\\
 and $\mathcal{E}_i(P_Y)=\emptyset$ for $i\geq 2$.




Note that $\mathcal{E}(P)=\cup_{i\ge
  0}\mathcal{E}_i(P)$ is finite up-to renaming of the variables since
  for every $i\geq 1$, every term $M\in \mathcal{E}_i(P)$, $\pos(M)$ is included
in the (finite) set of positions occurring in terms of $\mathcal{M}_0$.

We can now define an over-approximation of the set of tests that may be applied  over the secret.
\[\begin{array}{ll}
\mathcal{M}^{\x}_{t}(P) = &
\bigl\{T\in\mathcal{M}_t(P)\mid T=\x\text{ or }\exists p\in\posv(T)
\text{ s.t. }D_1(\dots D_n)\!=\!f_{d}(T,p)\neq\perp,\\
& \hspace*{-1.2cm} \exists E\in\mathcal{E}(P), \exists i\text{ s.t. }
D_i=\pi^i(\decg(\z,K)),  E=\encg(U,K,R) \text{ and } \y\in \nf{D_i(E)} \bigr\}
\end{array}
\]
For example,
$\mathcal{M}^{\x}_t(P_Y)=\set{\pi_1(\pi_2(\pi_2(\dec(\pi_1(z_a),k_{as}))))}$.

\begin{definition}\label{def:no_test_on_s}
We say that a well-formed process $P$ w.r.t.~$\x$  \emph{does not test
  over} $\x$
if the following conditions are satisfied:
\begin{enumerate}
 \item for all $E\in\mathcal{E}(P)$, for all $D=D_1(\dots D_n)\in\mathcal{D}_o(P)$, if
$D_i=\pi^i(\decg(\z),K)$ and $E=\encg(U,K,R)$ and $\y\in\fn(\nf{D_i(E)})$ then $i=1$ and $\bb{E}\not
<_{st}D_1$,
 \item if $[T=T']$, $[T'=T]$, $[\scheck(T,T',K)=\ok]$ or $[\scheck(T',T,K)=\ok]$ is a test of $P$ and $T\in\mathcal{M}^{\x}_{t}(P)$
 then $T'$ is a restricted name different from $\x$. 
\end{enumerate}
\end{definition}
For example, $P_Y$ does not test over $\x$.
Note that $\mathcal{E}(P)$ can  be computed in polynomial time from $P$ and that whether $P$ \emph{does not
test over} $\x$ is decidable. We show in the next section that the first condition is sufficient  to ensure
that frames obtained from $P$ are extended well-formed. It ensures in particular that there are no destructors right
above $\x$. If some $D_i$ cancels some encryption in  some $E$ and $\y\in\fn(\nf{D_i(E)})$ then
all its
destructors should  reduce  in the normal form computa\-tion (otherwise some destructors (namely projections
from $D_i$) remain above $\y$). Also we have $i=1$
since otherwise a $D_i$ may have consumed the lowest encryption above $\y$, thus the other
decryption may
block, and again there would be destructors left above~$\y$.

The second condition requires that whenever an operand of a  test
$[T=T']$ is potentially dangerous (that is $T$ or $T'$ is in
$\mathcal{M}^{\x}_{t}(P)$) then the other operand should be a
restricted name.

\begin{example} 
A simple class of protocols that do not test on the secret is the one where in all messages sent by
the protocol, the secret occurs only in the second component of pairs, and the tests apply
only on the first component of pairs. For example, if for a protocol $P_3$ we have
$\mathcal{M}_o(P_3)=\set{\enc(\pair{n_a}{\x},k,r),\enc(\pair{n_a}{\pi_2(\dec(z,k)),k',r')}}$ and the test is
$[\pi_1(\dec(z',k'))=\pi_1(\dec(z'',k))]$ then there will be no test on $\x$. Moreover, this protocol also
satisfies the first condition and hence we obtain that $\x$ is strongly secret using  the main result of
this section.

We also  give examples of protocols not satisfying the two conditions of
Definition~\ref{def:no_test_on_s}. Consider first a protocol $P_1$ for which
$\mathcal{M}_o(P_1)=\{\enc(\pi_1(\dec(z,k)),k,r'), \enc(\x,k,r)\}$. $P_1$ does not satisfy the first
condition of the previous definition because the term $\enc(\pi_1(\x),k,r)$ (with a destructor right above
$\x$) could be obtained by sending the first message to the agent which constructs the second message. 

A second example of protocol not satisfying the conditions (this time the second one) is inspired from the
Otway-Rees protocol. Consider a protocol $P_2$ where the server waits for $A,\penc{N_a,A}{K_{as}}$, performs
a test on $A$ and then sends $\penc{N_a,K_{ab}}{K_{as}}$. Using a second session, the intruder is able to
transform the test that the server does on $A$ into a test on the secret. Formally,
$\mathcal{M}_o(P_2)=\set{\pair{a}{\enc(\pair{n_a}{a},k_{as},r)},
\enc(\pair{\pi_1(\dec(\pi_2(z),k_{as}))}{\x}),k_{as},r'}$ and
$\mathcal{M}_t(P_2)=\{\pi_1(z),\pi_2(\dec(\pi_2(z),k_{as}))\}$. Then $\pi_2(\dec(\pi_2(z),k_{as}))\in
\mathcal{M}^{\x}_t(P_2)$ but $\pi_1(z)$ is not a restricted name.

\end{example}

\subsection{Main result}

We are now ready to prove that \syntactic secrecy is actually equivalent to strong secrecy for protocols that
are well-formed and do not test over the secret.

\newcommand{\thactive}{
Let $P$ be well-formed process w.r.t.~a bound name $\x$ such that $P$ does not
test over $\x$. We have $\varphi\nvdash\x$ for any valid frame $\varphi$ w.r.t.~$P$ if and only if
$P[\subst{\x}{M}]\approx_l P[\subst{\x}{M'}]$, for all ground terms $M,M'$ public w.r.t.~$\bn(P)$.
}
\begin{thm}\label{th-active}
\thactive
\end{thm}

\newcommand{\thpasact}{Let $\varphi$ be an extended well-formed frame \text{w.r.t.} $\x$,
where $\x$ is a restricted name in $\varphi$. Then $\varphi\nvdash
\x$  if and only if  $\varphi[\subst{\x}{M}]\approx
\varphi[\subst{\x}{M'}]$ for all $M, M'$ closed public terms w.r.t.~$\varphi$.}

\newcommand{\lemmacond}
{Let $P$ be a well-formed process with  no test over $\x$ and $\varphi=\nu\nb.\sigma$ be a valid frame
w.r.t.~$P$ such that $\varphi\nvdash\x$. Consider the corresponding standard
frame $\nu\nb.\bb{\sigma}=\nu\nb.\set{\subst{y_i}{U_{i}}\mid 1\le i\le l}$. For every $i$ and every
occurrence $q_{\x}$ of $\x$ in $\nf{U_i}$,  we have $f_e(\nf{U_i},q_{\x})=E[\subst{\y}{W}]$ for
some $E\in\mathcal{E}(P)$ and some term $W$. In addition $\nu\nb.\nf{\sigma_i}$ is an extended well-formed
frame {w.r.t.} $\x$.}

\newcommand{\lemmatests}{
Let $P$ be a well-formed process with no test over $\x$, $\varphi = \nu\nb.\sigma$ a valid frame for $P$ such
that $\varphi\nvdash\x$, $\theta$ a public substitution and $M$ a public ground term. If $T_1=T_2$ is a test
in $P$, then $T_1\theta\sigma[\subst{\x}{M}]=_E T_2\theta\sigma[\subst{\x}{M}]$ implies $T_1\theta\sigma=_E
T_2\theta\sigma$.}

\proof
Consider first the simpler implication, that is strong secrecy implies \syntactic secrecy.
Suppose that there is a valid frame $\varphi$ w.r.t.~$P$ such that $\varphi\vdash\x$. Then, as for the
passive case, there are $M$ and $M'$ public ground terms such that
$\varphi[\subst{\x}{M}]\not\approx\varphi[\subst{\x}{M'}]$. Since $\varphi$ is a valid frame there is an
extended process $A$ such that $P\Rightarrow^*A$ and $\varphi=\varphi(A)$. Then clearly
$P[\subst{\x}{M}]\Rightarrow^*A[\subst{\x}{M}]$ and $P[\subst{\x}{M'}]\Rightarrow^*A[\subst{\x}{M'}]$. Thus
if $P[\subst{\x}{M}]\approx_lP[\subst{\x}{M'}]$ then $A[\subst{\x}{M}]\approx_lA[\subst{\x}{M'}]$ and
moreover $\varphi(A[\subst{\x}{M}])\approx\varphi(A[\subst{\x}{M'}])$.
Since $\varphi(A[\subst{x}{T}])=\varphi(A)[\subst{x}{T}]$ for any term
$T$, we get
$\varphi[\subst{\x}{M}]\approx\varphi[\subst{\x}{M'}]$, contradiction.
We deduce $P[\subst{\x}{M}]\not\approx_lP[\subst{\x}{M'}]$ and thus
$\x$ is not strongly secret in $P$.

The remaining of the section is devoted to the converse implication. Let $P$ be well-formed process w.r.t.~a
bound name $\x$ with no test over
$\x$ and assume that $\x$ is \syntactically secret in $P$. 
Let $M,M'$ be to public terms w.r.t.~$\bn(P)$. To prove that $P[\subst{\x}{M}]$ and $P[\subst{\x}{M'}]$ are
labeled bisimilar, we need to show that each move of $P[\subst{\x}{M}]$ can be matched by a move in
$P[\subst{\x}{M'}]$ such that the corresponding frames are bisimilar (and conversely). By hypothesis, $P$ is
\syntactically secret w.r.t.~$\x$ thus for any valid frame $\varphi$ w.r.t.~$P$, we have
$\varphi\nvdash\x$. In
order to apply our previous result in the passive setting (Theorem~\ref{th}), we need to show that all the
valid frames are well-formed. However, frames may now contain destructors in particular if the adversary
sends messages that contain destructors.
That is why we consider \emph{extended  well-formed frames}, defined
in Section~\ref{sec:proof:passif}.


\smallskip

Theorem~\ref{th} can easily be generalized to extended well-formed frames.
\begin{prop}\label{th2}
\thpasact
\end{prop}
The proof of Proposition~\ref{th2} is exactly the same as the proof of Theorem~\ref{th} except that it uses
Corollary~\ref{corol_synt} and Lemma~\ref{lemma_simulate_rev} instead of Lemmas~\ref{lemma_syntactic}
and~\ref{lemma_simulate} respectively.

The first step of the proof of Theorem~\ref{th-active} is to show that any frame produced by the protocol is
an extended well-formed frame. We actually prove directly a stronger result, crucial in the proof: the secret
$\x$ always occurs under an agent encryption and this encryption is an instance of a term in $\mathcal{E}(P)$.
This shows that $\mathcal{E}(P)$ is indeed an approximation of the cyphertexts that may contain the secret.
\begin{lem}\label{lemma_cond}
\lemmacond
\end{lem}

The lemma is proved in Appendix~\ref{app:lemmas}.  The proof uses an induction on $i$ and relies deeply on the
construction of $\mathcal{E}(P)$.

The second step of the proof consists in showing that any successful test in the process $P[\subst{\x}{M}]$
is also successful in $P$ and thus in $P[\subst{\x}{M'}]$.
\begin{lem}\label{lemma_tests}
\lemmatests
\end{lem}
This lemma is proved in Appendix~\ref{app:lemmas} by case analysis,
depending on whether $T_1,T_2\in\mathcal{M}^{\x}_t(P)$ and whether $\x$
occurs or not in $\fn(T_1\theta\sigma)$ and $\fn(T_2\theta\sigma)$.


Using Lemmas~\ref{lemma_cond} and~\ref{lemma_tests}, we are ready to
complete the proof of Theorem~\ref{th-active}, showing that 
$P[\subst{\x}{M}]$ and $P[\subst{\x}{M'}]$ are labeled bisimilar.

We consider the relation $\mathcal{R}$ between closed extended processes defined as follows:
$A\,\mathcal{R}\,B$ if there is an extended process $A_0$ and ground terms $M,M'$ public w.r.t.~$\bn(P)$
such that $P\Rightarrow^* A_0$, $A=A_0[\subst{\x}{M}]$ and $B=A_0[\subst{\x}{M'}]$.

We show that $\mathcal{R}$ satisfies the three points of the definition of labeled bisimilarity. Suppose
$A\,\mathcal{R}\,B$, that is $A_0[\subst{\x}{M}]\,\mathcal{R}\,A_0[\subst{\x}{M'}]$ for some $A_0,M,M'$ as
above.

\begin{enumerate}
 \item Let us show that $\varphi(A_0[\subst{\x}{M}])\approx
 \varphi(A_0[\subst{\x}{M'}])$.
 We know that $\varphi(A_0)$ is a
valid frame w.r.t. $P$ (from the
definition of $\mathcal{R}$), hence $\varphi(A_0)\nvdash \x$ (from the hypothesis). Let
$\varphi'\equiv\varphi(A_0)$ having only ground and normalised terms (take for example
$\varphi'=\nf{\bb{\varphi(A)}}$, where $\bb{\varphi(A)}$ is the standard frame w.r.t.~$A$). Then, by
Lemma~\ref{lemma_cond}, we have that $\varphi'$ is an extended well-formed frame. We can then use
Proposition~\ref{th2} to obtain that $\varphi(A_0[\subst{\x}{M}])\approx \varphi(A_0[\subst{\x}{M'}])$.

 \item Let us show that if $A_0[\subst{\x}{M}]\rightarrow A'$ then $A'\equiv A'_0[\subst{\x}{M}]$, $A_0[\subst{\x}{M'}]
 \rightarrow A'_0[\subst{\x}{M'}]$ and
$A'_0[\subst{\x}{M}]\,\mathcal{R}\,A'_0[\subst{\x}{M'}]$, for some
 $A'_0$.
 We distinguish two cases, according
to whether the transition
rule was the \regle{COMM} rule or one of the \regle{THEN} and \regle{ELSE} rules:
\begin{itemize}
 \item if the \regle{COMM} rule was used then $A_0[\subst{\x}{M}]\equiv C[\subst{\x}{M}]\big[\out{c}{z}.
 Q[\subst{\x}{M}]|c(z).R[\subst{\x}{M}]\big]$, where $C$ is
an evaluation context and $A'=C[\subst{\x}{M}]\big[Q[\subst{\x}{M}]|R[\subst{\x}{M}]\big]$. Then
$A_0\equiv C[\out{c}{z}.Q|c(z).R]$. Take
$A'_0=C[Q|R]$. We have that $P\Rightarrow^*
A'_0$ and thus, by definition of $\mathcal{R}$, we have that
$A'_0[\subst{\x}{M}]\,\mathcal{R}\, A'_0[\subst{\x}{M'}]$.

 \item otherwise, $A_0[\subst{\x}{M}]\equiv C[\subst{\x}{M}]\big[\text{if}\ T'[\subst{\x}{M}]=T''[\subst{\x}{M}]\
\text{then}\ Q[\subst{\x}{M}]\ \text{else}\ R[\subst{\x}{M}]\big]$. Then
 $A_0\equiv C[\text{if}\ T'=T''\ \text{then}\ Q\ \text{else}\ R]$. From Lemma
 \ref{lemma_frame} we know that $T'=T'_0\theta\sigma$ and $T''=T''_0\theta\sigma$, where
$T'_0=T''_0$ is a test in $P$ and $\nu\nb.\sigma\equiv\varphi(A_0)$ is the standard frame {w.r.t.}
$A_0$. Take $A'_0=C[Q]$ if $T'_0\theta\sigma=_E T''_0\theta\sigma$ and
$A'_0=C[R]$ otherwise. From Lemma \ref{lemma_tests} we have that $T'_0\theta\sigma=_E
 T''_0\theta\sigma$ if and only if $T'_0\theta\sigma[\subst{\x}{M}]=_E T''_0\theta\sigma[\subst{\x}{M}]$. Hence
$A_0[\subst{\x}{M}]\rightarrow A'_0[\subst{\x}{M}]$,
 $A_0[\subst{\x}{M'}]\rightarrow A'_0[\subst{\x}{M'}]$ and $A_0\rightarrow A'_0$. We conclude
$A'_0[\subst{\x}{M}]\,\mathcal{R}\,
A'_0[\subst{\x}{M'}]$ from the definition of $\mathcal{R}$.
\end{itemize}

 \item Let us show that if $A_0[\subst{\x}{M}]\lab A'$ and $\fv(\alpha)\subseteq\dom(\varphi(A_0[\subst{\x}{M}]))$ and
$\bn(\alpha)\cap\fn(A_0[\subst{\x}{M'}])=\emptyset$
then $A'\equiv A'_0[\subst{\x}{M}]$, $A_0[\subst{\x}{M'}]\lab A'_0[\subst{\x}{M'}]$ and
$A'_0[\subst{\x}{M}]\,\mathcal{R}\, A'_0[\subst{\x}{M'}]$, for some $A'_0$.
Depending on the form of $\alpha$, we consider the following cases:
\begin{itemize}
 \item $\alpha=c(T)$. Suppose $A_0[\subst{\x}{M}]\equiv C[\subst{\x}{M}]\big[c(z).Q[\subst{\x}{M}]\big]$. Then
take $A'_0=C[Q\lbrace\subst{z}{T} \rbrace]$.
 \item $\alpha=\out{c}{u}$. Suppose $A_0[\subst{\x}{M}]\equiv
C[\subst{\x}{M}]\big[\out{c}{u}.Q[\subst{\x}{M}]\big]$. Then take
$A'_0=C[Q]$.
 \item $\alpha=\nu u.\out{c}{u}$. Suppose $A_0[\subst{\x}{M}]\equiv C[\subst{\x}{M}]\big[\nu
u.A_1[\subst{\x}{M}]\big]$, where
$A_1[\subst{\x}{M}]\stackrel{\out{c}{u}}{\longrightarrow}A'_1[\subst{\x}{M}]$. Then take $A'_0=C[A_1]$.
\end{itemize}
\end{enumerate}

The above discussion proves that $\mathcal{R}\subseteq\ \approx_l$.
Since we have $P[\subst{\x}{M}]\,\mathcal{R}\,P[\subst{\x}{M'}]$ it
follows that $P[\subst{\x}{M}]\approx_lP[\subst{\x}{M'}]$.\qed

\section{Application to some cryptographic protocols}

We apply our result to three protocols (Yahalom, Needham-Schroeder with symmetric keys and
Wide-Mouthed-Frog), known to preserve the usual syntactic secrecy
property.
Since all these three protocols satisfy our hypotheses, we directly
deduce that they preserve the strong secrecy property.

\subsection{Yahalom}
We have seen in Section~\ref{sec-hyp} that $P_{Y}$ is a well-formed process w.r.t.~$k_{ab}$ and does not
test over $k_{ab}$. Applying Theorem~\ref{th-active}, if $P_Y$ preserves the \syntactic secrecy of $k_{ab}$,
we
can deduce that the Yahalom protocol preserves the strong secrecy of
$k_{ab}$ that is
\[P_{Y}[\subst{k_{ab}}{M}]\approx_l P_{Y}[\subst{k_{ab}}{M'}]\]
for any public terms $M,M'$ w.r.t.~$\bn(P_{Y})$. We did not formally prove that the Yahalom protocol
preserves
the \syntactic secrecy of $k_{ab}$ but this was done with several tools in slightly different settings
(e.g.~\cite{BozgaLP03,Paulson01}).


In what follows, for sake of simplicity, we may omit the symbol $\langle,\rangle$ for pairing. In that case, we assume a
right priority that is $a,b,c = \langle\langle a,b\rangle,c\rangle$.

\subsection{Needham-Schroeder symmetric key protocol}
The Needham-Schroeder symmetric key protocol~\cite{NS78} is described below:
\[
\begin{array}[c]{rl}
A\Rightarrow S: & A,B,N_a \\
S\Rightarrow A: & \penc{N_a,B,K_{ab},\penc{K_{ab},A}{K_{bs}}}{K_{as}} \\
A\Rightarrow B: & \penc{K_{ab},A}{K_{bs}}\\
\end{array}
\]

The target secret is $K_{ab}$.
The protocol is modeled by the following process:
$$P_{\NS}=\nu k_{as}.\nu k_{bs}.\,(!A)\,|\,(!c(z_b))\,|\,(! \nu k.S(k))\,|\,\nu k_{ab}.S(k_{ab})$$
where
 \[\begin{array}[c]{rcl}
A& =& \nu
n_a.\out{c}{a,b,n_a}.c(z_a).[\pi_1(\dec(z_a,k_{as}))=n_a].\\&&\quad \quad
[\pi_1(\pi_2(\dec(z_a,k_{as})))=b].
\out{c}{\pi_2(\pi_2(\pi_2(\dec(z_a,k_{as}))))}
\\
S(x)& =& c(z_s).\nu
r,r'.\out{c}{\enc(\langle\pi_2(\pi_2(z_s)),\pi_1(\pi_2(z_s)),
k_{ab},\\&&\quad \quad
\enc(\langle x,\pi_1(z_s)\rangle,k_{bs},r' )\rangle,k_{as},r)}
\end{array}\]
Note that other processes should be added to considered corrupted agents or roles $A,B$ and $S$ talking to
other agents but this would not really change the following sets of messages.

The output messages are:
\[\mathcal{M}_o = \left\{\begin{array}{l}
a,b,n_a\\
\pi_2(\pi_2(\pi_2(\dec(z_a,k_{as}))))\\
\enc(\langle \pi_2(\pi_2(z_s)),\pi_1(\pi_2(z_s)),\\
 k_{ab},\enc(\langle k_{ab},\pi_1(z_s)\rangle,k_{bs},r' )\rangle,k_{as},r)
\end{array}\right\}
\]

The tests are: \comment{$\mathcal{M}_t = $}
\[\left\{\begin{array}{l}
\pi_1(\dec(z_a,k_{as}))=n_a\\
\pi_1(\pi_2(\dec(z_a,k_{as})))=b\\
\end{array}\right\}
\]


We define $\max \bb{\mathcal{E}_i}=\{\bb{e}\mid e\in\mathcal{E}_i\}$
in order to increase readability, and since it is easy to deduce
$\bb{\mathcal{E}_i}$ from $\max \bb{\mathcal{E}_i}$.

$$\mathcal{D}_o=\set{\pi_2(\pi_2(\pi_2(\dec(z,k_{as}))))}$$
\[\begin{array}{l}
\mathcal{E}_0=\{\enc(\pair{z_1}{\pair{z_2}{\pair{\y}{z_3}}},k_{as},r),
\enc(\pair{\y}{z_4},k_{bs},r')\}
\end{array}\]
$$\max
\bb{\mathcal{E}}_0=\set{\pi_1(\pi_2(\pi_2(\dec(z,k_{as})))),\pi_1(\dec(z,k_{bs}))}$$
$$\mathcal{D}_o\cap\bb{\mathcal{E}}_0=\emptyset$$
$$\mathcal{M}^{k_{ab}}_t=\emptyset$$

We deduce that $P_{\NS}$ is a well-formed process w.r.t.~$k_{ab}$, that
 does not test over~$k_{ab}$. Applying Theorem~\ref{th-active}
 and since the Needham-Schroeder symmetric key protocol
is known to preserve \syntactic secrecy of $k_{ab}$, we deduce that the protocol preserves strong secrecy of
$k_{ab}$
that is
\[P_{\NS}[\subst{k_{ab}}{M}]\approx_l P_{\NS}[\subst{k_{ab}}{M'}]\] for
any public terms $M,M'$ w.r.t.~$\bn(P_{\NS})$.

\subsection{Wide Mouthed Frog Protocol (modified)}
We consider a modified version of the Wide Mouthed Frog
Protocol~\cite{ban89}, where timestamps are replaced by nonces.
\[
\begin{array}[c]{rl}
A\Rightarrow B: & N_a\\
B\Rightarrow S: & \penc{N_a,A,K_{ab}}{K_{bs}} \\
S\Rightarrow A: & \penc{N_a,B,K_{ab}}{K_{as}}
\end{array}
\]

The target secret is $K_{ab}$.
The protocol is modeled by the following process:
$$P_{\WMF}=\nu k_{as}.\nu k_{bs}.\,(!A)\,|\,(!S)\,|\,(! \nu k.B(k))\,|\,\nu k_{ab}. B(k_{ab})$$
where
 \[\begin{array}[c]{rcl}
A&=&\nu n_a.\out{c}{n_a}.c(z_a).[\pi_1(\dec(z_a,k_{as}))=n_a]
\\
B(x)&=&c(z_b).\nu r.\out{c}{\enc(\langle z_b,a,x\rangle,k_{bs},r)}
\\
S& = &c(z_s).[\pi_1(\pi_2(\dec(z_s,k_{bs})))=a].\\& & \qquad\qquad
\nu
r'.\out{c}{\enc(\langle\pi_1(\dec(z_s,k_{bs})),b,\pi_2(\pi_2(\dec(z_s,k_{bs})))\rangle,k_{as},r')}
\end{array}\]
Note that other processes should be added to considered corrupted agents or roles $A,B$ and $S$ talking to
other agents but again, this would not really change the following sets of messages.

The output messages are:
\[\mathcal{M}_o = \left\{\begin{array}{l}
n_a\\
\enc(\langle z_b,a,k_{ab}\rangle,k_{bs},r)\\
\enc(\langle\pi_1(\dec(z_s,k_{bs})),b,\\
\pi_2(\pi_2(\dec(z_s,k_{bs})))\rangle,k_{as},r')
\end{array}\right\}
\]

The tests are:
\[\left\{\begin{array}{l}
\pi_1(\dec(z_a,k_{as}))=n_a\\
\pi_1(\pi_2(\dec(z_s,k_{bs})))=a
\end{array}\right\}
\]

$$\mathcal{D}_o=\set{\pi_1(\dec(z,k_{bs})),\pi_2(\pi_2(\dec(z,k_{bs})))}$$
$$\mathcal{E}_0=\set{\enc(\pair{z_1}{\pair{z_2}{\y},k_{bs},r)}}$$
$$\max\bb{\mathcal{E}}_0=\set{\pi_2(\pi_2(\dec(z,k_{bs})))}$$
$$\mathcal{E}_1=\set{\enc(\pair{z_1}{\pair{z_2}{\y},k_{as},r)}}$$
$$\max\bb{\mathcal{E}}_1=\set{\pi_2(\pi_2(\dec(z,k_{as})))}$$
$$\mathcal{D}_o\cap\bb{\mathcal{E}}_1=\emptyset$$
$$\mathcal{M}^{k_{ab}}_t=\emptyset$$

We obtain similarly that $P_{\WMF}$ is a well-formed process w.r.t.~$k_{ab}$, that
 does not test over~$k_{ab}$. Applying Theorem~\ref{th-active}
 and since the Wide Mouthed Frog protocol is known to
preserve \syntactic secrecy of $k_{ab}$, we deduce that the protocol preserves strong secrecy of
$k_{ab}$
that is
\[P_{\WMF}[\subst{k_{ab}}{M}]\approx_l P_{\WMF}[\subst{k_{ab}}{M'}]\] for
any public terms $M,M'$ w.r.t.~$\bn(P_{\WMF})$.

\section{Conclusion}

In recent years many automatic tools have been developed for verifying security protocols. 
The overwhelming majority of them address reachability-based properties such as 
\syntactic secrecy. On the other hand some important security notions such as  strong secrecy rely on provable 
equivalences between systems. Typically the impossibility of guessing a vote or a password is 
commonly expressed that way. 
Hence in order to widen the scope of the current protocol analysis tools, in the present paper 
we have shown how  \syntactic secrecy actually implies strong secrecy in both
passive and active setting under some conditions, motivated by counterexamples.
In particular such a result cannot  hold for deterministic encryption 
and we had to assume that it is \emph{probabilistic}. 


As future works, we plan to further investigate the active case by trying to relax our conditions. There are several
possible directions. First, we may consider specific classes of
protocols by restricting the syntax (for instance considering protocols without pairs such as in ~\cite{amadio02,HuttelS05}) 
to see whether it is possible to refine our results in this setting. 
Second, we may relax  the requirement that processes cannot test over the secret
by requiring instead that the two branches of the test are
indistinguishable. This is the case for example when a test is
followed in each branch by other tests that will never succeed when
the first one is really applied to a secret data. 
This would require to consider more complex over-approximations of the
set of sent messages. In particular, in the definition of the set
$\mathcal{E}$, we would have to consider trees instead of simply
paths potentially leading to the secret.


\bibliographystyle{plain}
\bibliography{biblio}


\clearpage
\appendix





\section{Proof of Lemma~\ref{lemma_simulate_rev}}\label{proof_lemma_simulate_rev}

\begin{lem}
\lemmasimulaterev
\end{lem}

\proof
Let $U,V,M$ be terms with $U$ and $M$ public w.r.t.~$\varphi$, $M$ being closed and in normal form such that
$U\sigma[\subst{\x}{M}]\rightarrow V$, as in the statement of the lemma. Let $L\rightarrow
R\in\mathcal{R}_E$ be the rule that was applied in the above reduction and let $p$ be the position at which
it was applied, \textit{i.e.}~$U\sigma[\subst{\x}{M}]|_p=L\theta$.
Since $M$ is in normal form, $p\in\pos(U\sigma)$.

Assume that there is a substitution $\theta_0$ such that $U\sigma|_p=L\theta_0$. This will be proved in
the Claim below. It follows that $U\sigma$ is reducible.
If $p\not\in\posnv(U)$ then there is a term of $\ran(\sigma)$ which is
reducible. This contradicts the fact that $\varphi$ is an extended-well formed frame (since all terms in
such a frame should be in normal form). Hence we have that $p\in\posnv(U)$.
Let $T=U|_p$. We have $T\sigma[\subst{\x}{M}]=L\theta$ and $T\sigma=L\theta_0$.

For our equational theory $E$, $R$ is either a constant (\emph{i.e.}~$\ok$) or a
variable.  If $R$ is a constant then we take $V'=U[R]_p$ and
$\sigma'=\sigma$. It is easy to verify that the conditions of the
lemma are satisfied in this case.

Suppose now that $R$ is a variable $z_0$. Then, consider the\footnote{For our equational theory there is
exactly one occurrence of $z_0$ in $L$. \comment{ but generally, it is sufficient to consider an arbitrary
position, since $L|_q$ is the same term for different $q$'s (this comes from the existence of $\theta_0$
such that $T\sigma=L\theta_0$).}} position $q$ of $z_0$ in $L$. This position $q$ is also in
$L\theta_0$, that is in $T\sigma$.
Hence the two following possibilities may occur:
\begin{enumerate}
\item If $q\in\posnv(T)$, that is there is no $y\in\dom(\sigma)$ above
$z_0$, then we consider $V'=U[T|_q]_p$ and $\sigma'=\sigma$. In this
case also, it is easy to verify that the conditions of the
lemma are satisfied.

\item If $q\notin\posnv(T)$, that is there is some
$y\in\dom(\sigma)$ above $z_0$, then we consider $V'=U[y']_p$ and
$\sigma'=\sigma\cup\{R\theta_0/y'\}$, where $y'$ is a new variable (i.e.\ $y'\notin\dom(\sigma)$).
The term $V'$ is clearly public {w.r.t.} $\varphi'$.
Since $T\sigma=_E R\theta_0$, $\varphi\vdash R\theta_0$. This shows that $\varphi\vdash W$ if
and only if $\varphi'\vdash W$ for any term $W$.

\smallskip
\noindent We have $V'\sigma'=(U[y']_p)\sigma'=U\sigma'[y'\sigma']_p=U\sigma[R\theta_0]_p$. Hence
$U\sigma\rightarrow
V'\sigma'$.

\smallskip
\noindent From $T\sigma=L\theta_0$ and $T\sigma[\subst{\x}{M}]=L\theta$
\comment{and since $x\notin\var(L)$,} we deduce that
$z\theta_0[\subst{\x}{M}]=z\theta$ for all $z\in\var(L)$, hence
$R\theta_0[\subst{\x}{M}]=R\theta$. Thus
$V'\sigma'[\subst{\x}{M}]=(U\sigma[\subst{\x}{M}])[R\theta]_p=V$.

\smallskip
\noindent Since there is some $y\in\dom(\varphi)$ above $z_0$,
$R\theta_0 = z_0\theta$ is a subterm of a term of $\sigma$.
Then $R\theta_0$ is in normal form since all
the terms in $\ran(\sigma)$ are in normal form. Also all agent encryptions in $\varphi'$ are probabilistic.
Suppose that
there is an occurrence of $\x$ in $R\theta_0$ such that there is no encryption plaintext-above it (in
$R\theta_0$). In
this case we have that all the function symbols above this occurrence in $R\theta_0$ are $\langle\rangle$ or
$\sign$. Thus $\x$ is deducible from $\varphi'$ and hence from $\varphi$, which represents a
contradiction with the hypothesis. Hence there is an encryption plaintext-above any occurrence of $\x$ in
$R\theta_0$.
 All this proves that $\varphi'$ is also an extended well-formed frame.
\end{enumerate}

\noindent {\bf Claim}: Let us now prove that there exists  $\theta_0$ such that $U\sigma|_p=L\theta_0$.
Assume by contradiction that it is not the case. Then at least one of the following cases occurs:
\begin{enumerate}
 \item there is a position in $L$ which is not a position in $U\sigma|_p$;
 \item there is a variable $z$ in $L$ having at least two occurrences, say at positions $p_1, p_2$, for which
$(U\sigma|_p)|_{p_1}\neq (U\sigma|_p)|_{p_2}$.
\end{enumerate}

Let us examine in detail the two cases:
\begin{enumerate}
 \item Consider a minimal  position $q'$ (w.r.t.~the prefix order) in $L$ which is not a position in
$U\sigma|_p$.
Then $q'=q\cdot 1$ with $q$ position of $U\sigma|_p$ and there is an
$\x$ at position $q$ in $U\sigma|_p$ (since such minimal positions in $L$ must be
positions in $U\sigma[\subst{\x}{M}]|_p$, but not in $U\sigma|_p$). Also $q\neq\epsilon$
(\textit{i.e.} it does not correspond to the head of $L$) since otherwise $M$ would not be in normal form.
By examining all rules in $\mathcal{R}_E$, we observe  that at least one of the conditions in the definition
of extended well-formed frames is not satisfied.  For example, if $L\rightarrow R$ is the rule
$\pi_1(\pair{z_1}{z_2})\rightarrow z_1$ then $q=1$. Then either $\pi_1(y)$ is the subterm at
position $p$ in $U$ and $y\sigma=\x$ (impossible case since $\x$ would be deducible), either $\pi_1(\x)$ is
the subterm at position $p$ in $U\sigma$ and this subterm is also a subterm of a term of $\sigma$ (again an
impossible case because there are no destructors right above $\x$ in term of an extended well-formed frame).
If $L\rightarrow R$ is the rule $\deca(\enca(z_1,\pub(z_2),z_3),\priv(z_2))\rightarrow $ then $q$ might be
$1$ or $1\pdot 2$. The case $q=1$ is similar with the previous one. If $q=1\pdot 2$ then we have a term in
$\sigma$ having $\enca(W,\x)$ as subterm for some $W$ (otherwise $\x$ would be deducible). But this again
contradicts the definition of extended well-formed frames. The analysis for the other rules is similar.

 \item Let $T_1=(U\sigma|_p)|_{p_1}$ and $T_2=(U\sigma|_p)|_{p_2}$. We have $T_1\neq T_2$, but
$T_1[\subst{\x}{M}] = T_2[\subst{\x}{M}]$. Consider an arbitrary position $q_{\x}$ of $\x$ in $T_1$.
Since $U$ is public, there is a variable $y\in\var(U)$ at position say $p_y$ such that $p_y\le
p\pdot p_1\pdot q_{\x}$.
Consider the lowest agent encryption $q_{\enc}$ plaintext-above $q_{\x}$ in $U\sigma$.
It occurs in $y\sigma$ according to the definition of extended well-formed frames. Suppose that
$p\pdot p_1> q_\enc$. The function symbols between $q_\enc$ and $p\pdot p_1$ must be
$\langle\rangle$ or $\sign$. But this doesn't hold for none of rules in $\mathcal{R}_E$. Hence there is an
agent encryption plaintext-above $q_\x$ in $T_1$. The same argument applies to $T_2$.
We can thus use Point \ref{corol_condsub} of Corollary~\ref{corol_synt} to $T_1$ and $T_2$ and obtain a
contradiction, that is $T_1=T_2$.
\end{enumerate}
We have seen that the two cases lead to contradictions. So there is $\theta_0$ such that
$U\sigma|_p=L\theta_0$.\qed

\section{Proof of Lemma~\ref{lemma_frame}}\label{proof_lemma_frame}

\begin{lem}
\lemmaframe
\end{lem}

\proof
We provide an inductive and constructive proof. We reason by induction on the number of reductions in
$P\Rightarrow^* A$.

The base case is evident.

Assume that $P\Rightarrow^l A_k$ and that there are $l$, $B_l$ and $\theta$ as in the statement of the
lemma. Suppose that $A_l \Rightarrow A_{l+1}$ and consider the reduction rule that was used:
\begin{itemize}
 \item If it is an internal reduction then, since static equivalence is closed by structural
equivalence and by internal reduction (see Lemma 1 in~\cite{AbadiFournet01}), it is sufficient to consider
as searched values the same as for $A_l$.

\comment{we have that $\varphi(A_l)\equiv \varphi(A_{l+1})$ and
the messages does not change.}

\comment{Suppose, for example, that the \regle{COMM} was used, that is,
$A_l\equiv C[\out{c}{x}.P \mid c(x).Q]$ and $A_{l+1}\equiv C[P \mid Q]$.}

 \item If it is a labeled reduction then we prove the following property: $\alpha\neq \out{c}{x}$ (for any
$a$ and $x$) and  there is an extended process $B_{l+1}=\varphi(B_{l+1})|P_{l+1}$ such that $B_{l+1}\equiv
A_{l+1}$ and
\begin{itemize}
\item if $\alpha=\nu x.\out{c}{x}$ then $P_{l+1}=P_l$ and $\varphi(B_{l+1})=\nu\nb.\sigma_{k+1}$, where
$\sigma_{k+1}=\sigma_k\cup\set{\subst{x}{M_l}}$ and $M_l$ is an output in $P_l$.

\item if $\alpha=c(M)$ then $\varphi(B_{l+1})=\varphi(B_l)$ and for every message (an operand of a test or an
output) $M_{l+1}$ in $P_{l+1}$ there is a message (an operand of a
test or an output, respectively) $M_l$ in $P_l$, such that
$M_{l+1}=M_l\theta'\sigma_k$, for some substitution $\theta'$ public
{w.r.t.} $\nu\nb$.

\item if $\alpha=\out{c}{n}$ or $\alpha=\nu n.\out{c}{n}$ then $P_{l+1}=P_l$, and
$\varphi(B_{l+1})=\varphi(B_l)$ or  $\varphi(B_{l+1})=\nu \set{\nb}\backslash \set{n}.\sigma_k$,
respectively.
\end{itemize}

It is easy to see that this property is sufficient to prove the inductive step.

The property can be verified, by showing, using induction on the shape of the derivation tree, that for any
extended processes $A',A'',B'$ such that $A'\lab A''$, $A'\equiv B'$, $B'=\nu\nb.\sigma | Q$  there is $B''$
such that $A''\equiv B''$ and $B'=\nu\nb'.\sigma' | Q'$ where
\begin{itemize}
 \item if $\alpha=c(M)$ then $\nb'=\nb$, $\sigma'=\sigma$ and $N''=N'\set{\subst{x}{M}}$ for each term
$N''$ of $B''$ where $N'$ is the corresponding term in $B'$ and $c(x)$ is an input in $B'$;
 \item if $\alpha=\nu x.\out{c}{x}$ then $Q'=Q$, $\nb'=\nb$, and $\sigma'=\sigma\cup\set{\subst{x}{M}}$
 where $\out{c}{M}$ is an input in $B'$;
 \item if $\alpha=\out{c}{x}$, $\alpha=\out{c}{n}$ or $\alpha=\nu n.\out{c}{n}$ then $\nb'=\nb$ for the
 first two cases, and $\set{\nb'}=\set{\nb}\backslash\set{n}$ for the third one, $\sigma'=\sigma$ and
 $Q'=Q$.
\end{itemize}
\end{itemize}

\comment{ Suppose the last used rule was
\begin{itemize}
 \item \regle{STRUCT}
 \item
\end{itemize}

\textbf{Idea} We use the \regle{ALIAS} rule only for making possible the application of \regle{COMM} and
\regle{OUT-ATOM}, we never use the \regle{REWRITE} rule (except maybe for the \regle{THEN}, depending on
how it is defined) and we always use the \regle{SUBST} rule. ...

Observations:

Since we don't use the \regle{ALIAS} rule it means that the terms from $P$ "are preserved" in $A$: we
didn't replace $M=N$ from $P$ with $\nu z,z'.(\set{\subst{z}{M},\subst{z'}{N}}|(z=z'))$ in $A$; nor
$\out{c}{m}$ with $\nu z.\set{\subst{z}{m}}|\out{c}{z}$.

Since we don't use the \regle{REWRITE} rule we have syntactic equality when saying $M=M_0\theta\sigma$.

If $A\equiv\varphi(A)|(\nu\overline{n}.\out{c}{m}.P|Q)$ then $A\equiv\nu
y.(\set{\subst{y}{m}}|\varphi(A)|(\nu\overline{n}.\out{c}{y}.P|Q))$. }\qed

\section{Proof of lemmas \ref{lemma_cond} and \ref{lemma_tests}}\label{app:lemmas}

In what follows we usually simply write $\mathcal{M}$, $\mathcal{M}_t$, $\mathcal{M}_o$,
$\mathcal{D}_o$, $\mathcal{E}$ instead of respectively $\mathcal{M}(P)$, $\mathcal{M}_t(P)$,
$\mathcal{M}_o(P)$, $\mathcal{D}_o(P)$, $\mathcal{E}(P)$, \textit{etc.}

We also define the partial subtraction function
$-:\mathbb{N}_+^*\times\mathbb{N}_+^*\rightarrow\mathbb{N}_+^*$ as follows: $p-q=r$ if $p=q\pdot r$ and
$p-q=\perp$ otherwise.

Let $U$ and $V$ be  two terms. We define $\pos(U,V)=\set{p\in \pos(U)\mid U|_p=V}$.

Observe that for the rewriting system corresponding to  equational theory $E$, there is at most one rule
that can be applied and for each rule $R\rightarrow L$, there is exactly
one occurrence of $R$ in $L$. 

We denote by $U\rightarrow^q V$ the reduction $U\rightarrow V$ such that $U|_q=L\theta$ and
$V=U[R\theta]_q$, where $q$ is a position in $U$,  $L\rightarrow R$ is a rule in $\mathcal{R}_E$, and
$\theta$ is a substitution. Let $p$ be a position in $U$. 
We define a 
partial function $\nfp_1(U,p,q)$ that computes, when $U\rightarrow^q V$, the \emph{position after one
rewriting} of a function symbol at position $p$ in $U$.
In particular, if $\nfp_1(U,p,q)\neq \perp$ then
$U|_p=V|_{\nfp_1(U,p,q)}$.
Formally, we define the function
$\nfp_1\colon\mathcal{T}\times\mathbb{N}^*_+\times\mathbb{N}^*_+\rightarrow
\mathbb{N}^*_+$ as follows:
\[\nfp_1(U,p,q)=\left\{
\begin{array}{ll}
p', & \text{ if }U\rightarrow^q V\\
\perp, & \text{ otherwise,}
\end{array}
\right.\] where
\[p'=\left\{
\begin{array}{ll}
p, &  \text{ if }p\not\ge q,\\
\perp, &  \text{ if }p\ge q\ \wedge\ p\not\ge q\pdot q_r,\\
q\pdot (p-q\pdot q_r), &  \text{ if }p\ge q\pdot q_r,
\end{array}
\right.\]\index{sf}and $L\rightarrow R$ is the rule that was applied and $q_r$ is the position of $R$ in
$L$.

Similarly, the function $\nfp(U,p)$ computes the \emph{position after
rewriting} in $\nf{U}$. 
The function
$\nfp\colon\mathcal{T}\times\mathbb{N}_+^*\hookrightarrow\mathbb{N}_+^*$
is formally defined by $\nfp(U,p)=p_k$ where
$U\rightarrow^{q_1}\dots\rightarrow^{q_k} U_k$, $U_k=\nf{U}$, $p_i=\nfp_1(U,p_{i-1},q_i)$, for $1\le i \le
k$ and $p_0=p$. 
Due to the particular form of our equational theory, the choice of the
rewriting steps does not change the final value of $p_k$ thus 
the definition is correct. 

The function
$\nfpinv(U,p)$ is the inverse function: to a position $p$ in $\nf{U}$ it associates the corresponding position
in $U$, that is, $\nfpinv\colon\mathcal{T}\times\mathbb{N}_+^*\hookrightarrow\mathbb{N}_+^*$, $\nfpinv(U,p)=p'$ if and only
if $\nfp(U,p')=p$.

We say that a function symbol at position $p$ is \emph{consumed in $V$ {w.r.t.} the reduction}
$U\rightarrow^q V$ if $\nfp_1(U,p,q)$ is undefined. Similarly, we say that  a function symbol at position $p$ is 
\emph{consumed in $\nf{U}$ {w.r.t.} the normal form} $\nf{U}$ if
$\nfp(U,p)$ is undefined. We say simply that an occurrence is consumed in some term when it is clear from the
context which definition is used.


\begin{lem}
\lemmacond
\end{lem}

\proof
We write the standard frame $\bb{\sigma}$ as in the statement of Lemma~\ref{lemma_frame}, that is
$U_i=M_i\theta_i\sigma_{i-1}$ for all $1\le i\le l$ with $M_i$ an output in $P$, $\theta_i$ a public
substitution w.r.t $\x$ and $\sigma_i=\sigma_{i-1}\cup\set{\subst{y_i}{U_i}}$, $\sigma_0$ being the empty
substitution. We reason by induction on $i$.

\smallskip
Base case: $i=1$. We have that $U_1=M_1\theta_1$. Then $\nf{U_1}=M_1(\nf{\theta_1})$ since there are no
destructors in the output $M_1$. Hence any position $q_{\x}$ of $\x$ is in fact a position in $M_1$ since $\x$ cannot
appear in $\theta_1$ because $\x$ is restricted and $\theta$ is a public substitution. There must an
encryption above $q_{\x}$ in $M_1$ (that is a position $q_{\enc}\pdot 1\le q_{\x}$), since otherwise $\x$
would be deducible (the same argument as in Lemma~\ref{lemmadeduc} applies). Then the result follows
immediately from the definition of $\mathcal{E}_0$ (take $W=\x$) and the properties of well-formed processes.

\smallskip
Inductive step. Let $p_{\x}=\nfpinv(U_i,q_{\x})$.

If $p_{\x}\in\pos(M_i)$ then, as in the previous
paragraph, $f_e(\nf{U_i},q_{\x})[\subst{\x}{\y}]\in\mathcal{E}_0$.

\smallskip
Otherwise, since $\theta_i$ is public, $p_{\x}\notin\pos(M_i\theta)$. It follows that there are
$z\in\var(M_i)$ and $y_{i_1}\in\var(M_i\theta_i)$ at positions $p_z$ and $p_{y_1}$ respectively, such that
$p_z\le p_{y_1}\le p_{\x}$ and $1\le i_1 \le i-1$. Let $p^1_{\x}=p_{\x}-p_{y_1}$ and
$q^1_{\x}=\nfp(U_{i_1},p^1_{\x})$. By induction hypothesis, $\sigma_{i-1}$ is an extended well-formed frame
and $f_e(\nf{U_{i_1}}, q^1_{\x})=E[\subst{\y}{W}]$ with $E\in\mathcal{E}_l$, for some term
$W$ and some $l\ge 0$. It follows from the definition of extended well-formed frames that in
$y_1\sigma_{i_1}$ there is an encryption above $q^1_{\x}$, that is
$q^1_{\enc}=\max\set{\,q\in\pos(\nf{U_{i_1}})\mid q<q^1_{\x}\ \wedge\
\head{(U_{i_1}\downarrow)|_{q}}\!=\encg\,}$ exists. Let $p^1_{\enc}=\nfpinv(U_{i_1},q^1_{\enc})$.

\smallskip
If $p_{y_1}\pdot p^1_{\enc}$ is not consumed in $\nf{U_i}$ then
$\nfp(U_i,p_{y_1}\pdot p^1_{\enc})$ is the lowest encryption in $\nf{U_i}$ above $q^1_{\x}$ (since it
corresponds to $q^1_{\enc}$). It follows that $f_e(\nf{U_i},q_{\x})=f_e(\nf{U_{i_1}}, q^1_{\x})$.

\smallskip
Otherwise, that is if $p_{y_1}\pdot p^1_{\enc}$ is consumed in~$\nf{U_i}$, consider the occurrence of
$\decg$ in $U_i$, say $p_{\dec}$, that consumes it. Since $p^1_{\enc}$ is not consumed w.r.t.~$\nf{U_{i_1}}$
it follows that $p_{\dec}\in\pos(M_{i}\theta_i)$, and all encryptions above $p^1_{\enc}$ in
$U_{i_1}$ are consumed in $\nf{U_i}$. If $p_{\dec}$ is in $z\theta_{i}$ (that is,
$p_{\dec}\notin\posnv(M_i)$) then all encryptions above $p^1_{\enc}$ in $U_{i_1}$ are consumed by decryptions
that are in $z\theta_{i}$. This means that in $\nf{(z\theta_{i}\sigma_{i-1})}$ there is no encryption above
$\x$ and thus $\varphi\vdash\x$.
Hence $p_{\dec}$ is in $M_{i}$ (that is, $p_{\dec}\in\posnv(M_i)$).

Let $U, V, K, K'$ and $R$ be terms such that $\decg(U,K)=U_i|_{p_{\dec}}$ and
$\encg(V,K',R)=U_i|_{p_{y_1}\pdot p^1_{\enc}}=U_{i_1}|_{p^1_{\enc}}$. We have that $K=_EK'$ since $p_{\dec}$
consumes $p_{y_1}\pdot p^1_{\enc}$.
We then have $\decg(U,K)\rightarrow^* \decg(\encg(V,K,R),K)\rightarrow^* \nf{V}$.

Let $(D,p)=f_{dp}(M_i,p_z)$ and write it as $D=D_1(\dots D_n)$ where $D_j=\pi^j(\decg(\z,K_j))$ with $1 \leq
j \leq n$ and consider $D_k$ such that the decryption $p_{\dec}$ is that of $D_k$. Clearly
$\y\in\fn(\nf{D_j(E)})$.
From the first condition of processes that do not test over $\x$ we have that $j=1$ and
$\bb{E}\not<_{st}D_1$. Since $p_{\dec}$ consumes $p_{y_1}\pdot
p^1_{\enc}$, above $p_{\dec}$ in $D_1$ there are only projections, below $\encg$ in $E$ there are
only pairs and $\bb{E}\not<_{st}D_1$ it follows that $D_1\le_{st}\bb{E}$. Hence $D_1\in\mathcal{\bb{E}}_l$.

Suppose that there is no encryption above $p_{\dec}$ in $M_i$. Then since $D_1$ is consumed and
above $D_1$ in $M_i$ there are only pairs or signatures, it follows that $\x$ is deducible
from $\sigma_i$ (more exactly from $\nf{U_i}$). Thus there is at least one encryption above $p_{\dec}$ in
$M_i$. Let $(M',p_{\enc})=f_{ep}(M_{i},p_z)$. Then $M'[\y]_p\in\mathcal{E}_{l+1}$.


Since $p_{\enc}$ is not consumed in $\nf{U_i}$ and in $M'$ all function symbols above $p$ are
not destructors we have that $f_e(U_i,p_{\x})$ $\rightarrow^*$ $(M'[\y]_p)[\y\rightarrow
D_1(f_e(\encg(V,K',R),p'_{\x}))]$ where $p'_{\x}=p^1_{\x}-p^1_{\enc}$. Hence
$f_e(\nf{U_i},q_{\x})=(M'[\y]_p)[\subst{\y}{W'}]$, where $W'=\nf{D_1(f_e(\encg(V,K',R),p'_{\x}))}$.
That is we have the first part of the lemma.

\smallskip
In order to prove that $\nf{\sigma}$ is an extended well-formed frame we just need show that $M'[\y]_p$ and
$W'$ contain only pairs and signatures (except for the head of $M'[\y]_p$ which is an
encryption); obviously all agent encryptions are probabilistic encryption, either by the definition of
well-formed process or by induction hypothesis. From the definition of $M'$ all function
symbols (except for the head) in $M'[\y]_p$ are pairs and signatures. And since
$\sigma_{i_1}$ is an extended well-formed frame and the term $W'$ is a subterm of
$f_e(\encg(\nf{V},K',R),q'_{\x})$ which (except for the head) contains only pairs as function symbols and
signatures by definition of $f_e$.\qed

\medskip
\noindent\textbf{Claim.}
Let $P$ be a well-formed process with no test over $\x$,
$\varphi=\nu\nb.\sigma$ be a valid frame w.r.t.~$P$ such that
$\varphi\nvdash\x$, $T\in\mathcal{M}_t(P)$ be an operand of a
test and $\theta$ be a public substitution. If
$T\notin\mathcal{M}^{\x}_t$ then for any occurrence $q_{\x}$ of $\x$ in
$\nf{(T\theta\sigma)}$ there is an encryption $q_{\enc}$ plaintext-above it such that
this encryption is an agent encryption w.r.t.~$\nb\!\setminus\!\set{\x}$,
is a probabilistic encryption w.r.t.~$\ran(\sigma)$ and
$\head{{(T\theta\sigma)}\downarrow|_q}\in\{\langle\rangle,\sign\}$,
for all positions $q$ with $q_{\enc}<q<q_{\x}$.

\proof
Suppose that $T\notin\mathcal{M}^{\x}_t$ and consider an occurrence $q_{\x}$ of
$\x$ in $\nf{(T\theta\sigma)}$. Hence $T$ is not
ground and denote by $z$ the variable of $T$ and by $p_z$ its position.
Let $T_z=\nf{(z\theta\sigma)}$.

Let $\bb{\sigma}=\set{\subst{y_1}{U_1},\dots,\subst{y_l}{U_l}}$ be the standard frame w.r.t.~$A$
(where $\varphi=\varphi(A)$ for some extended process $A$).
Let $p_{\x}=\nfpinv(T\theta\bb{\sigma},q_{\x})$. Let $y_{i}$ be the variable of $z\theta$ on the path
to $p_{\x}$ at position say $p_{y}$, with $1\le i\le l$.
Applying Lemma~\ref{lemma_cond} to $U_i$ we obtain that $f_e(\nf{U_{i}}, q_{\x})=E[\subst{\y}{W}]$ with
$E\in\mathcal{E}(P)$, for some term $W$. Consider the lowest encryption $q_{\enc}$ in
$\nf{U_i}$ above $q'_{\x}$, where $q'_{\x}$ is the position in $\nf{U_i}$ of $q_{\x}$.

Suppose that this encryption is consumed. Then it must be consumed by a $\decg$ from $T$ since otherwise
$\x$ would be deducible. It follows that there is $1\le j\le l$ such that $D_j=\pi^j(\dec(\z,K))$, where
$f_d(T,p_z)=D_1(\dots D_n)$, $E=\enc(U,K,R)$ and $\y\in \nf{D_i(E)}$ for some terms $U$, $K$ and $R$. Thus
$T\in\mathcal{M}^{\x}_t$, but this contradicts the hypothesis. Hence $q_{\enc}$ is not consumed in
$\nf{(T\theta\sigma)}$. Since $\nu\nb.\nf{\sigma}$ is an extended well-formed frame (again from Lemma~\ref{lemma_cond})
then the encryption $q_{\enc}$ clearly satisfies the hypothesis.\qed

\begin{lem}
\lemmatests
\end{lem}
\proof

$T_1\theta\sigma[\subst{\x}{M}]=_E T_2\theta\sigma[\subst{\x}{M}]$ rewrites in
$\nf{(T_1\theta\sigma[\subst{\x}{M}])}=\nf{(T_2\theta\sigma[\subst{\x}{M}])}$. Since
the rewrite system $\mathcal{R}_E$ is convergent, it follows that
$\nf{(\nf{(T_1\theta\sigma)}\,[\subst{\x}{M}])}=\nf{(\nf{(T_2\theta\sigma)}\,[\subst{\x}{M}])}$.

Suppose first that $T_1,T_2\not\in\mathcal{M}^{\x}_t$. Then from the claim above any occurrence of
$\x$ there are no destructors, hence $\nf{(T_1\theta\sigma)}[\subst{\x}{M}]$
is already in normal form. The same thing holds for $T_2$.
Thus $\nf{(T_1\theta\sigma)}[\subst{\x}{M}] = \nf{(T_2\theta\sigma)}[\subst{\x}{M}]$.
The previous claim also ensures that in
$\nf{(T_1\theta\sigma)}$ and $\nf{(T_2\theta\sigma)}$ there is an agent probabilistic encryption
above each occurrence of $\x$. Hence we can apply Lemma~\ref{lemma_synt} and obtain that
$\nf{(T_1\theta\sigma)} = \nf{(T_2\theta\sigma)}$, that is ${T_1\theta\sigma} =_E {T_2\theta\sigma}$.

Suppose now that $T_1\in\mathcal{M}^{\x}_t$. Then $T_2=n$ where $n$ is a restricted name. The name
$n$ is a subterm
of $\nf{(T_1\theta\sigma[\subst{s}{M}])}$ appearing at a position $p$ in $T_1\theta\sigma[\subst{s}{M}]$. Since $M$ is public,
while $T_2$ is restricted it follows $n$ is not a subterm of $M$, that is there
is no occurrence $q_{\x}$ of $\x$ in $T_1\theta\sigma$ such that $q_{\x}\le p$. Then
$\nf{(\nf{(T_1\theta\sigma)}[\subst{s}{M}])} = \nf{(T_1\theta\sigma)}[\subst{s}{M}]$. Hence
$\nf{(T_1\theta\sigma)} = n$.

If the test is $\scheck(T,T',K)=\ok$ then $T\theta\sigma[\subst{\x}{M}]=_E\retr(T')\theta\sigma[\subst{\x}{M}]$.
Applying the lemma for the test $T=_E\retr(T')$ we obtain that
$T\theta\sigma=_E\retr(T')\theta\sigma$. Since the keys are ground then it follows that $\scheck(T,T',K)\theta\sigma=_E\ok$.\qed

\end{document}